\documentclass[11pt]{article}

\usepackage[final]{acl}

\usepackage{times}
\usepackage{latexsym}

\usepackage[T1]{fontenc}

\usepackage[utf8]{inputenc}

\usepackage{microtype}

\usepackage{inconsolata}

\usepackage{graphicx}

\usepackage{hyperref}
\usepackage{url}
\usepackage{xcolor}

\usepackage{subcaption} 

\usepackage{booktabs}
\usepackage[table]{xcolor}
\usepackage{multirow}
\usepackage{multicol}
\usepackage{pifont}

\usepackage{paralist}
\usepackage{enumitem}

\usepackage{wrapfig}
\usepackage{amsmath}
\usepackage{arydshln}

\usepackage{cleveref}
\crefname{figure}{Figure}{Figures}
\Crefname{figure}{Figure}{Figures}
\crefname{table}{Table}{Tables}
\Crefname{table}{Table}{Tables}
\crefname{section}{Section}{Sections}
\Crefname{section}{Section}{Sections}
\crefname{appendix}{Appendix}{Appendices}
\Crefname{appendix}{Appendix}{Appendices}

\definecolor{myblue}{RGB}{175,208,240}
\definecolor{myyellow}{RGB}{251,240,182}
\definecolor{magenta}{RGB}{217,47,138}

\usepackage{siunitx}
\usepackage{listings}
\title{\raisebox{-0.328em}{\includegraphics[height=1.3em]{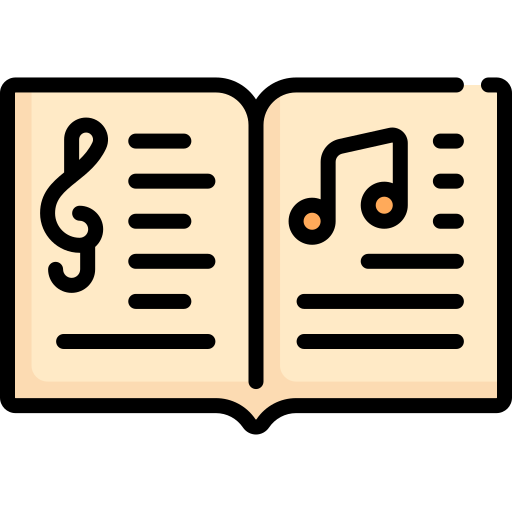}}\hspace{0.35em}Musical Score Understanding Benchmark: Evaluating Large Language Models' Comprehension of Complete Musical Scores}

\usepackage{fontawesome5}
\usepackage{wasysym}
\newcommand{\symfootmark}[2]{\textsuperscript{#2}}

\newcommand{\affilicon}[1]{\textsuperscript{\raisebox{-0.22ex}{\includegraphics[height=1.6ex]{#1}}}\kern-0.05em}
\newcommand{\affiliconshift}[2]{\textsuperscript{\raisebox{#2}{\includegraphics[height=1.6ex]{#1}}}\kern-0.01em}
\newcommand{\affiliconsized}[2]{\textsuperscript{\raisebox{-0.27ex}{\includegraphics[height=1.4ex]{#1}}}\kern0.05em}
\newcommand{\ccom}{\affilicon{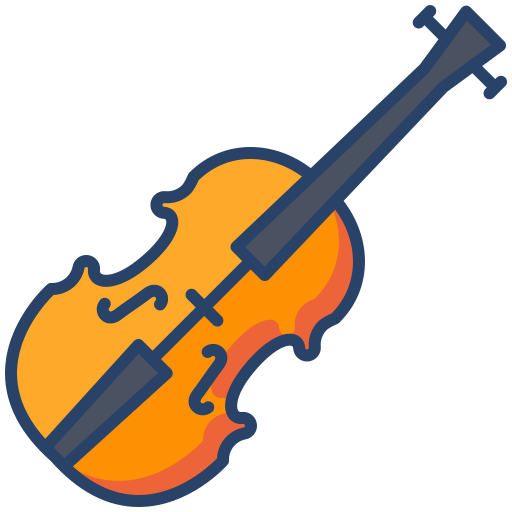}}
\newcommand{\imperial}{\affiliconshift{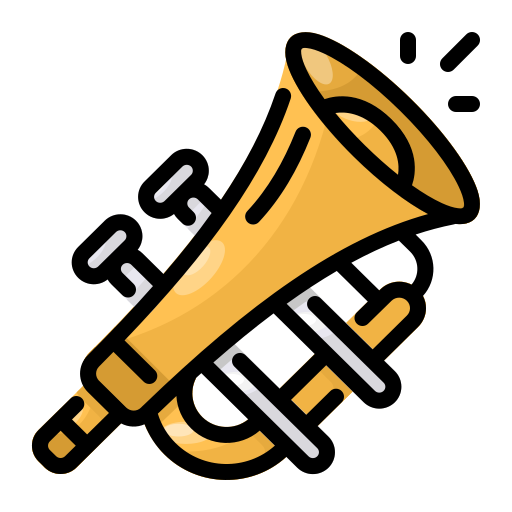}{-0.35ex}}
\newcommand{\tsinghua}{\affiliconsized{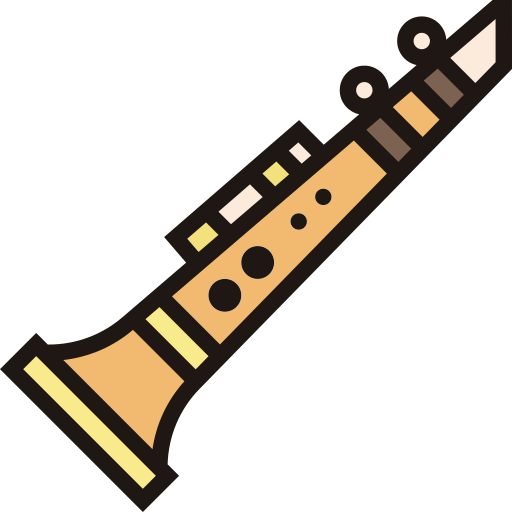}{1.38ex}}

\author{Congren Dai\ccom\imperial\tsinghua\symfootmark{fn:eq}{*}\symfootmark{fn:affil}{†},
Yue Yang\imperial\symfootmark{fn:eq}{*},
Krinos Li\imperial\symfootmark{fn:eq}{*}, Huichi Zhou\imperial, \\
\textbf{Shijie Liang\ccom, Bo Zhang\ccom, Enyang Liu\ccom, Ge Jin\ccom, Hongran An\ccom\tsinghua\symfootmark{fn:affil}{†}, Haosen Zhang\imperial,} \\
\textbf{Peiyuan Jing\imperial, Kinhei Lee\imperial, Zhenxuan Zhang\imperial, Xiaobing Li\ccom, \textnormal{and} Maosong Sun\ccom\tsinghua\symfootmark{fn:affil}{†}\symfootmark{fn:corr}{‡}} \\
\ccom Central Conservatory of Music \quad
\imperial Imperial College London \quad
\tsinghua Tsinghua University \\
\small{
    \texttt{congren.dai@\{mail.ccom.edu.cn,imperial.ac.uk\}\quad sms@tsinghua.edu.cn}
}
}

\usepackage{xspace}
\newcommand{\BenchShort}{MSU-Bench\xspace}
\newcommand{\BenchLong}{Musical Score Understanding Benchmark\xspace}

\begin{document}
\maketitle
{
\renewcommand{\thefootnote}{\fnsymbol{footnote}}
\footnotetext[1]{\hypertarget{fn:eq}{}Equal contribution.}
\footnotetext[2]{\hypertarget{fn:affil}{}Also affiliated with the Key Laboratory of Music and Brain Science, Central Conservatory of Music, Ministry of Education, China; the Laboratory of Music AI, Central Conservatory of Music, Laboratory of Philosophy and Social Sciences, Ministry of Education, China; and Natural Language Processing and Computational Social Science Lab (THUNLP), Tsinghua University, China.}
\footnotetext[3]{\hypertarget{fn:corr}{}Corresponding author.}
}

\begin{abstract}
Understanding complete musical scores entails integrated reasoning over pitch, rhythm, harmony, and large-scale structure, yet the ability of Large Language Models and Vision--Language Models to interpret full musical notation remains insufficiently examined.
We introduce \BenchLong (\BenchShort), a human-curated benchmark for score-level musical understanding across textual (ABC notation) and visual (PDF) modalities.
\BenchShort contains 1,800 generative question--answer pairs from works by Bach, Beethoven, Chopin, Debussy, and others, organised into four levels of increasing difficulty, ranging from onset information to texture and form.
Evaluations of more than fifteen state-of-the-art models, in both zero-shot and fine-tuned settings, reveal pronounced modality gaps, unstable level-wise performance, and challenges in maintaining multilevel correctness.
Fine-tuning substantially improves results across modalities while preserving general knowledge, positioning \BenchShort as a robust foundation for future research in multimodal reasoning. The benchmark and code are available at \url{https://github.com/Congren-Dai/MSU-Bench}.
\end{abstract}

\section{Introduction}
\label{sec:intro}

Large Language Models (LLMs) and Vision--Language Models (VLMs) have recently exhibited strong capabilities in natural language understanding and generation, driving substantial advances across a broad range of Natural Language Processing tasks \citep{NEURIPS2020_1457c0d6, chowdhery2022palmscalinglanguagemodeling, openai2024gpt4technicalreport, openai2025gpt5}. In contrast, their ability to reason over complete musical scores remains underexplored. Existing benchmarks for musical score understanding are typically narrow in scope, concentrating on isolated fragments, short excerpts, or multiple-choice formulations, rather than supporting holistic reasoning over entire scores. Moreover, most prior work focuses on monophonic music, which consists of a single melodic line without harmonic or rhythmic accompaniment. Such settings fail to reflect the structural complexity and expressive richness required for open-ended, real-world musicological analysis.

\begin{figure*}[t]
    \centering
    \includegraphics[width=1\linewidth]{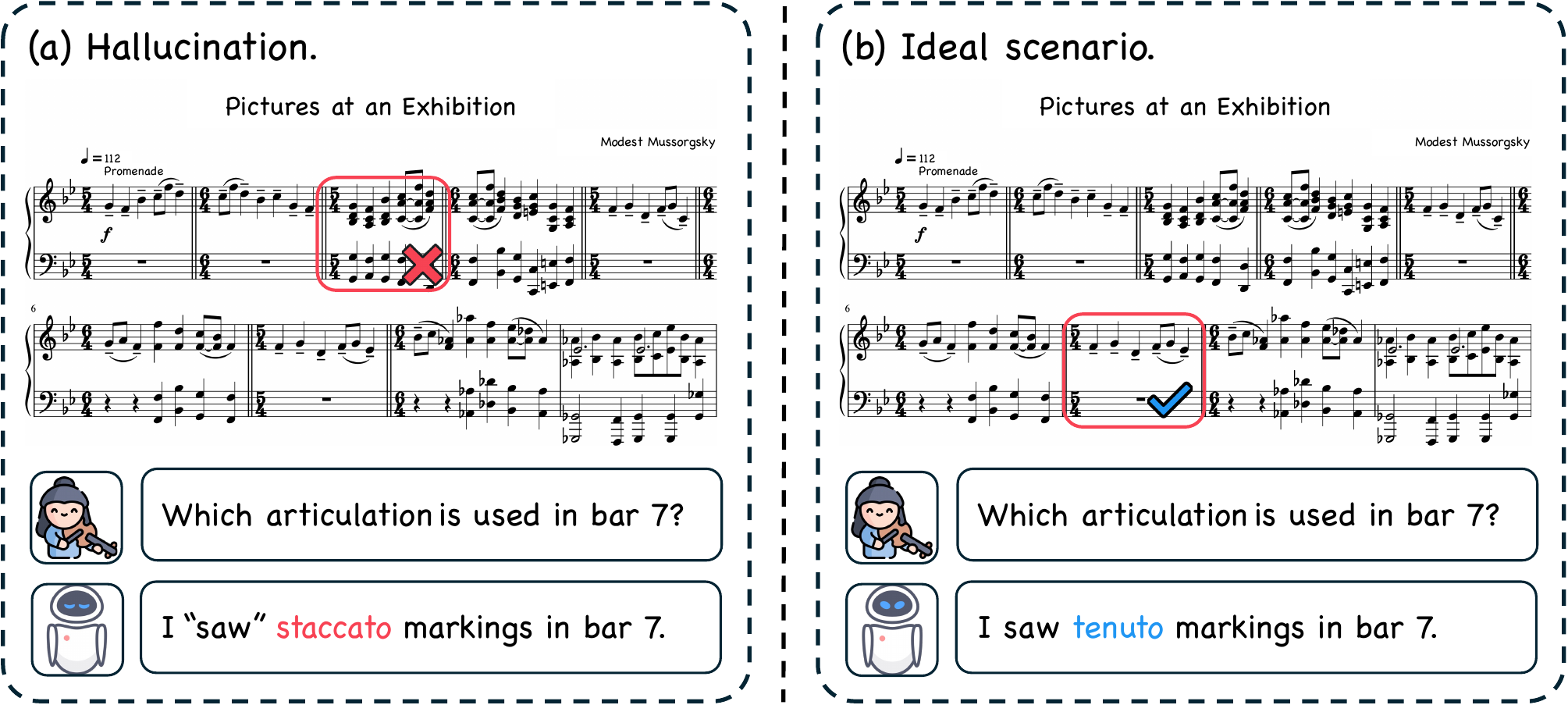}
    \caption{
        (a) \textbf{Hallucination.} When queried about specific score features in bars, VLMs often fabricate responses that are not grounded in the actual score. 
        (b) \textbf{Ideal scenario.} Models should accurately localise and analyse bars, thereby supporting reliable higher-level musicological reasoning.
    }
    \label{fig:problem}
\end{figure*}

When applied to complete scores, VLMs encounter two persistent challenges. The first is localisation: models frequently struggle to correctly identify bar positions, which is a prerequisite for answering higher-level questions related to harmony, texture, or form. For instance, when asked ``Which articulation is used in bar~7?'', models often misalign the bar index and consequently return incorrect markings (see \cref{fig:problem}a). The second challenge is hallucination, whereby models generate content that is not grounded in the score, often exacerbating errors introduced by incorrect localisation. Together, these issues lead to unreliable interpretations of complete scores and undermine confidence in model outputs relative to ideal, score-faithful answers (see \cref{fig:problem}b).

To measure these limitations, we curate \BenchLong (\BenchShort), a benchmark designed to evaluate the reasoning capabilities of LLMs and VLMs on complete musical scores, with particular emphasis on bar identification and higher-level musical understanding. The benchmark comprises 150 complete scores and 1,800 human-curated question--answer (QA) pairs, drawn primarily from representative textbook material used in undergraduate conservatory curricula. This educational motivation reflects the premise that models capable of answering such score-based questions could function as effective instructional assistants for music students. The benchmark is organised into four hierarchical levels of musical comprehension: Onset Information, Notation and Note, Chord and Harmony, and Texture and Form. These levels range from basic recognition of notational elements to advanced harmonic and structural analysis. By explicitly targeting localisation and hallucination while requiring reasoning over complete scores, \BenchShort provides a systematic foundation for evaluating textual and visual musical reasoning in contemporary models.

We evaluate LLMs using textual input encoded in ABC notation and VLMs using visual input in the form of PDF representations of complete scores. Our experiments demonstrate that the aforementioned challenges can be substantially mitigated by representing scores in ABC notation \citep{walshaw2011abcstandard}. In this work, we treat ABC as a structured upper-bound modality for symbolic-to-theory reasoning when reliable bar and voice structure is available, while PDF inputs remain the end-to-end setting for visual score understanding. ABC notation is a text-based symbolic format that explicitly encodes bar structure, pitch, rhythm, and articulation using human-readable characters, thereby offering a compact, LLM-friendly representation; by contrast, MusicXML preserves richer notational detail but introduces longer sequences and more serialisation choices. In this work, we make the following contributions: (1) we introduce MSU-Bench, a benchmark for evaluating LLMs and VLMs on complete musical scores, consisting of 1,800 human-curated generative QA pairs spanning four hierarchical levels of musical comprehension; (2) \BenchShort supports multimodal evaluation through textual question answering with ABC notation and visual question answering with PDF scores; (3) zero-shot evaluation on more than fifteen models reveals a pronounced gap between textual and visual modalities, unstable performance across comprehension levels, and limited robustness; (4) fine-tuning leads to performance gains in both modalities while preserving general knowledge; and (5) we observe that posing questions jointly yields better performance than presenting them sequentially, suggesting that current models can effectively leverage hierarchical reasoning.

\section{Related Work}
\begin{figure*}[t]
    \centering
    \includegraphics[width=1\linewidth]{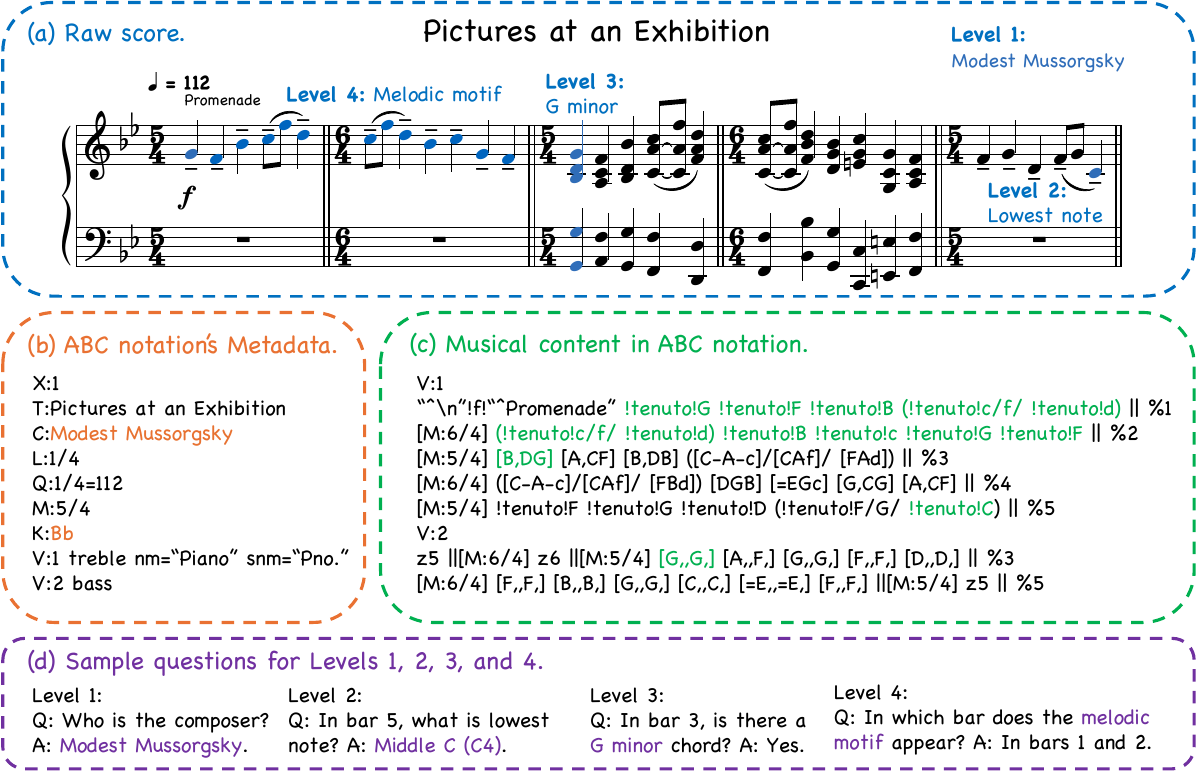}
    \caption{
        Illustration of multi-level understanding in MSU-Bench using Mussorgsky’s \textit{Pictures at an Exhibition}. Q indicates a question, while A indicates an answer.
        (a) Raw score excerpt with annotated tasks across four levels of difficulty. 
        (b) Metadata encoded in ABC notation. 
        (c) Musical content in ABC notation. 
        (d) Sample questions for each level, showing progression from foundational concepts to higher-level musical reasoning.}
    \label{fig:abc}
\end{figure*}

\textbf{Musical Score Representation.}
Musical score understanding constitutes a key task of Music Information Retrieval (MIR), aiming to analyse and interpret symbolic music representations in order to support downstream applications such as genre and style recognition \citep{simonetta2019multimodal}. 
Drawing on approaches in representation learning, earlier studies have frequently employed Optical Music Recognition (OMR) to convert scores into digital formats, such as MIDI \citep{moore1988dysfunctions}, MusicXML \citep{good2001musicxml}, and LilyPond \citep{nienhuys2003lilypond}, thereby facilitating the learning of embeddings that capture musical structure and semantics for these understanding tasks \citep{zeng2021musicbert, liang2020pirhdy, chou2021midibert}. 
On the other hand, musical notation systems such as ABC notation encode musical elements in a text format. This representation is typically written using ASCII characters \citep{gorn1963american}.
Its concise, highly compressed, and language-compatible format makes it particularly suited for integration with large language models, enabling symbolic music understanding and generation \citep{tang2025nota,wang2025notagen}.

\noindent
\textbf{QA Benchmarks for Score Understanding.} 
This area has attracted increasing interest in QA tasks for score understanding, which require more advanced forms of musical comprehension \citep{yue2024mmmu}. Notably, \mbox{MusicTheoryBench} \citep{yuan2024chatmusicianunderstandinggeneratingmusic} represents a systematic attempt to assess the competence of LLMs in music theory, evaluating performance across tasks that demand both music knowledge and reasoning. 
MusiXQA \citep{chen2025musixqaadvancingvisualmusic} evaluates VLMs in their ability to interpret musical scores represented as images. 
ZIQI-EVal \citep{li2024musicmaestromusicallychallenged} benchmarks LLMs on tasks of music comprehension and generation, with particular emphasis on their capacity to integrate contextual and cultural background knowledge. 
Furthermore, SSMR-Bench \citep{wang2025synthesizingsheetmusicproblems} introduces a synthetic data generation framework capable of producing both textual and visual question formats to support comprehensive evaluations of musical understanding.
Recently, WildScore \citep{mundada2025wildscorebenchmarkingmllmsinthewild} introduces an in-the-wild benchmark that combines real score images with musicological multiple-choice questions (MCQs) sourced from public forums.

\section{Benchmark Design}
\label{sec:degisn}

\subsection{Research Questions}
\label{sec:RQs}
\BenchShort aims to inspire future research on musical score understanding using LLMs and VLMs. In particular, it investigates the following research questions (RQs):

\begin{figure*}[t]
    \centering
    \includegraphics[width=1\linewidth]{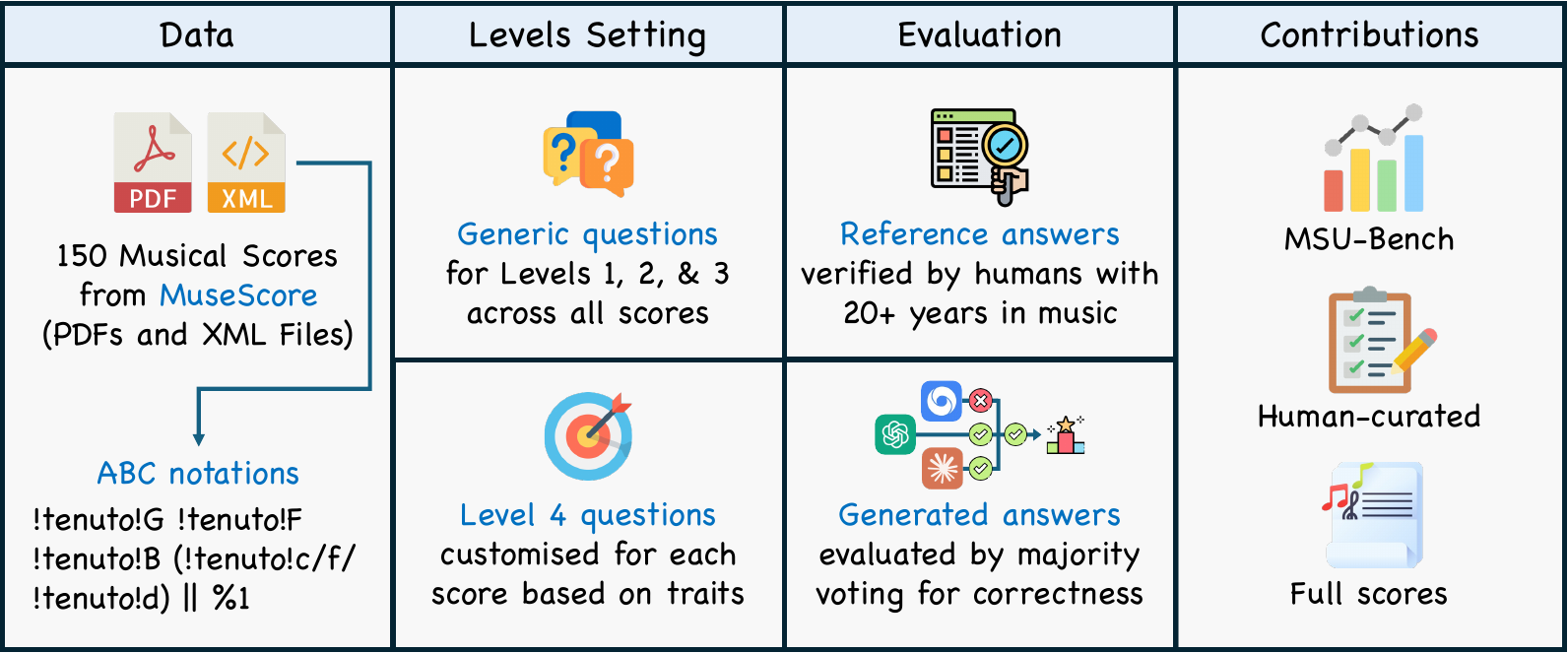}
    \caption{
        MSU-Bench data curation and evaluation framework. We collect 150 musical scores from MuseScore in PDF and MusicXML formats and convert them into ABC notation. Levels 1--3 use generic question templates applicable to all scores, while Level 4 questions are crafted for the unique musical features of each piece. We then verify all 1,800 questions manually. To evaluate whether generated answers match the reference answers, we employ an ``LLM-as-a-judge'' protocol in which LLMs independently determine whether each generated answer is correct or incorrect; the final decision is obtained through majority voting.}
    \label{fig:data}
\end{figure*}

\noindent
\textbf{\textcolor{magenta}{{RQ1: How accurately can a model identify onset-level musical metadata?}}}

\noindent
\textbf{Level 1 (Onset Information).} Level 1 evaluates a model’s ability to extract onset-level metadata from symbolic scores, forming the basis for subsequent analysis. This includes identity information such as composer, title, and instrumentation; notational elements such as key, clef, and time signature; performance indicators including tempo and expressive markings; and initial structural features such as the presence of an anacrusis.

\noindent
\textbf{\textcolor{magenta}{{RQ2: How correctly can a model interpret local notational and pitch-level features?}}}

\noindent
\textbf{Level 2 (Notation \& Note).} Level 2 shifts from global metadata to local, bar-level notation. It assesses the identification of pitch range, accidentals, rests, ornaments, articulations, dynamics, clef, key and time signature changes, tempo changes, and repeat signs, capturing localised details essential for texture and performance understanding.

\noindent
\textbf{\textcolor{magenta}{{RQ3: To what extent can a model accurately analyse harmonic structures in symbolic scores?}}}

\noindent
\textbf{Level 3 (Chord \& Harmony).} Level 3 addresses higher-order harmonic organisation, such as chord qualities and functions, inversions, voicing, spacing, and the handling of omitted or repeated notes. It further examines multi-bar progressions, cadential patterns, pedal points, ornamental harmonic devices, and the tracing of key and tonal changes from opening through modulation to tonal closure.

\noindent
\textbf{\textcolor{magenta}{{RQ4: To what extent can a model analyse textural and formal aspects of musical works?}}}

\noindent
\textbf{Level 4 (Texture \& Form).} Level 4 extends harmonic analysis to global texture and form, focusing on the identification and development of melodic motifs, thematic organisation, accompaniment types, scoring, and orchestration. It also encompasses broader formal categories and requires sensitivity to registral distribution across bars, voices, or instruments. Level 4 prompts ask about the most frequent texture or accompaniment type in a piece, the location of principal or secondary themes, where a melodic motif first appears, how that motif is developed or ornamented, and the main structural features of the motif.

\subsection{Case Study}
We present a case study illustrating the structure of Level 1--4 questions in \BenchShort, showing that ABC notation functions as a medium for musical understanding rather than a purely textual surrogate for the score, as illustrated in \cref{fig:abc}a.
ABC notation comprises two components, metadata (\cref{fig:abc}b) and musical content (\cref{fig:abc}c). Together, they encode both structural and performance-related information from Mussorgsky’s \textit{Pictures at an Exhibition}, providing sufficient symbolic detail to support questions across all four levels (\cref{fig:abc}d).

\noindent
\textbf{Metadata Information.}
The ABC header supports Level 1 questions. It specifies the title (\texttt{T:Pictures at an Exhibition}), composer (\texttt{C:Modest Mussorgsky}), default note length (\texttt{L:1/4}), tempo (\texttt{Q:1/4=112}), time signature (\texttt{M:5/4}), and key (\texttt{K:Bb}). Voice assignments are given by \texttt{V:1 treble nm=``Piano'' snm=``Pno.''} and \texttt{V:2 bass}, defining the treble and bass staves.

\noindent
\textbf{Musical Content.}
The musical body preserves bar-level structure and all information required for Levels 2--4. In \cref{fig:abc}c, bar indices are explicitly marked using \texttt{\%1}--\texttt{\%5}. Notation such as \texttt{!tenuto!C} identifies pitch-level detail within bar 5, while chord symbols \texttt{[B,DG]} and \texttt{[G,,G,]}, together with the key \texttt{K:Bb}, imply a G-minor harmony in bar 3, where commas indicate octave displacement below the reference register. Melodic structure is encoded directly, with the motif in bar 1 given as \texttt{!tenuto!G !tenuto!F !tenuto!B (!tenuto!c/f/ !tenuto!d)} and its transformation in bar 2 as \texttt{(!tenuto!c/f/ !tenuto!d) !tenuto!B !tenuto!c !tenuto!G !tenuto!F}, preserving phrase-level information essential for higher-level musical reasoning.

\begin{figure}
    \centering
    \includegraphics[width=\linewidth]{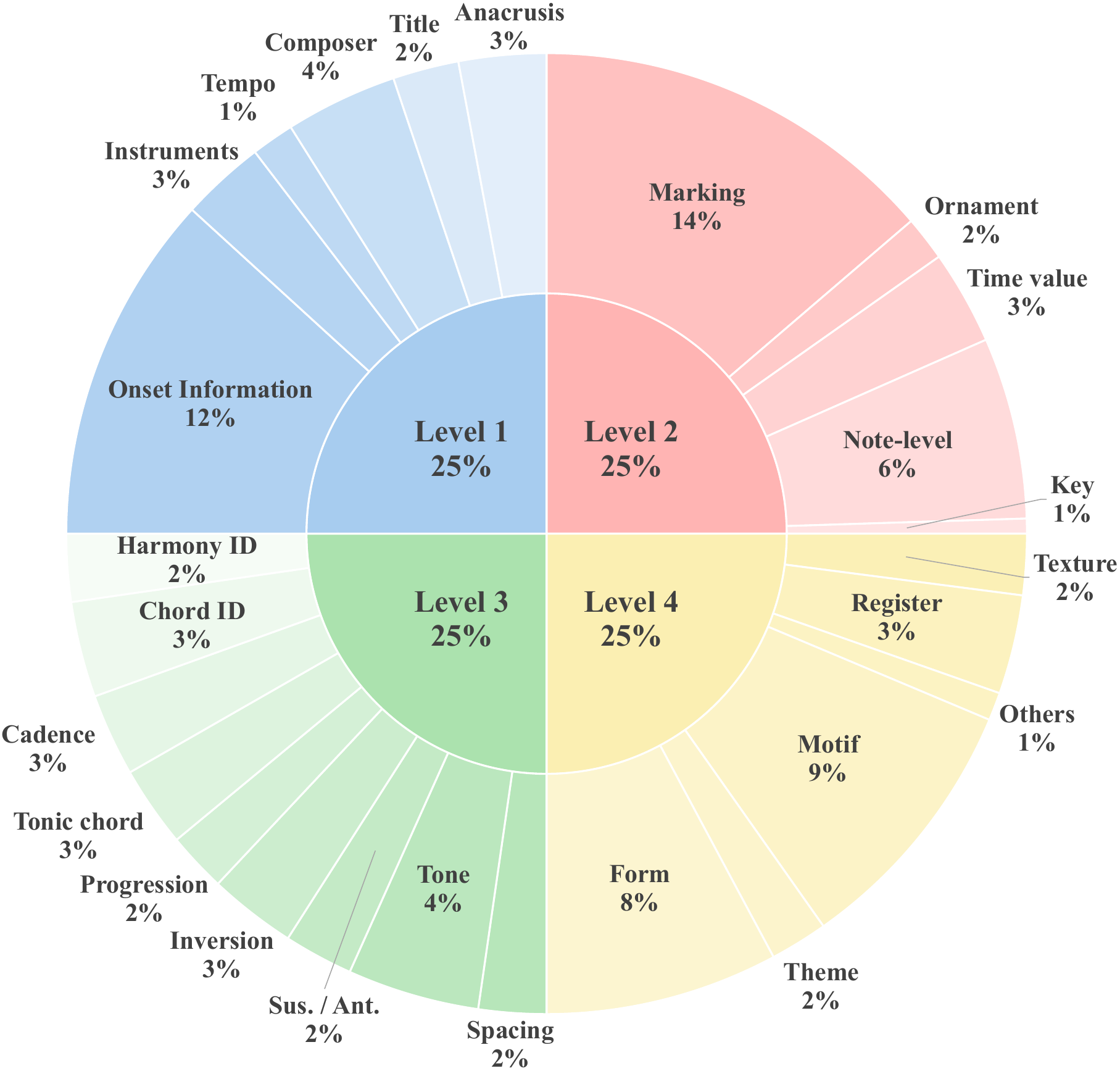}
    \caption{Distribution of Questions in \BenchShort.}
    \label{fig:level}
\end{figure}

\newcommand{\cmark}{\ding{51}} 
\newcommand{\xmark}{\ding{55}} 
\newcommand{\tmark}{$\triangle$} 

\begin{table*}[t]
\centering
\caption{Comparison of music-related QA benchmarks across multiple dimensions. A checkmark (\cmark) indicates the presence of a feature, a cross (\xmark) denotes its absence, and a triangle (\tmark) represents partial coverage.}

\resizebox{\textwidth}{!}{%
\begin{tabular}{@{}lcccccccc@{}}
\toprule
\multirow{2}{*}{\textbf{Benchmark}} &
  \multicolumn{2}{c}{\textbf{Modality}} &
  \multirow{2}{*}{\textbf{Sheet Music QA}} &
  \multirow{2}{*}{\textbf{Trainable}} &
  \multirow{2}{*}{\textbf{Homophony}} &
  \multirow{2}{*}{\textbf{QA Type}} &
  \multirow{2}{*}{\textbf{Quantity}} &
  \multirow{2}{*}{\textbf{Source}} \\ \cmidrule(lr){2-3}
 &
  \textbf{Textual} &
  \textbf{Visual} &
   &
   &
   &
   &
   &
   \\ \midrule
MMMU \citep{yue2024mmmu} &
  \xmark &
  \cmark &
  \cmark &
  \xmark &
  \cmark &
  \textbf{MCQs} &
  \textbf{369} &
  \textbf{Web} \\
MusiXQA \citep{chen2025musixqaadvancingvisualmusic} &
  \xmark &
  \cmark &
  \xmark &
  \cmark &
  \cmark &
  \textbf{Generative} &
  \textbf{1.3M} &
  \textbf{Synthetic} \\
ZIQI-Eval \citep{li2024musicmaestromusicallychallenged} &
  \cmark &
  \xmark &
  \xmark &
  \xmark &
  \cmark &
  \textbf{MCQs} &
  \textbf{14244} &
  \textbf{LLMs} \\
MusicTheoryBench \citep{yuan2024chatmusicianunderstandinggeneratingmusic} &
  \cmark &
  \xmark &
  \tmark &
  \xmark &
  \xmark &
  \textbf{MCQs} &
  \textbf{372} &
  \textbf{Human-labelled} \\
SSMR-Bench \citep{wang2025synthesizingsheetmusicproblems} &
  \cmark &
  \cmark &
  \cmark &
  \cmark &
  \xmark &
  \textbf{MCQs} &
  \textbf{3200} &
  \textbf{Synthetic} \\
WildScore \citep{mundada2025wildscorebenchmarkingmllmsinthewild} &
  \cmark &
  \cmark &
  \cmark &
  \cmark &
  \tmark &
  \textbf{MCQs} &
  \textbf{807} &
  \textbf{In-the-wild} \\
\BenchShort (Ours) &
  \cmark &
  \cmark &
  \cmark &
  \cmark &
  \cmark &
  \textbf{Generative} &
  \textbf{1800} &
  \textbf{Human-labelled} \\ \bottomrule
\end{tabular}%
}

\label{tab:compare}
\end{table*}

\subsection{Data Curation and Evaluation Process}
The data collection and model evaluation processes for \BenchShort are shown in \cref{fig:data}. The process begins with the selection of 150 scores from MuseScore. The complete list of scores included in \BenchShort is provided in \Cref{sec:score}. Our sampling unit is a score instance, which may correspond to a full score, a movement, or a self-contained excerpt, rather than a unique work title. For visual QA, the PDF of each score is employed, whereas for textual QA, the corresponding MusicXML file is converted into ABC notation. All PDFs originate from MuseScore source files and are exported directly from the corresponding project files using consistent default settings, such as page layout and rendering parameters, avoiding scan artefacts and preserving engraving clarity. We manually inspect the exported PDFs and ensure that bar numbers are visible at the beginning of each system or segment, so bar-indexed questions are not confounded by inconsistent bar-number availability. A comprehensive set of general questions is then developed and categorised into three levels of difficulty (Levels 1--3), designed to evaluate a broad range of musical concepts encompassing fundamental notational knowledge. In addition, score-specific questions are designed for Level 4. Representative examples of Levels 1--4 questions are provided in \Cref{sec:sample_question}. With the exception of Level 1, Levels 2--4 are intentionally designed to include bar localisation tasks.

For simpler scores, we formulate questions so that each level continues to probe the intended musical ability for that piece. When a concept is not meaningfully instantiated in a particular score, we use a musicologically appropriate equivalent within the same level definition rather than reducing the task to a trivially easier variant.

To ensure high-quality reference answers, all answers are first provided manually and then verified by human domain experts possessing over 20 years of professional musical experience (detailed in \Cref{sec:Instructions Given To Annotators,sec:Recruitment And Payment}). This procedure guarantees that \BenchShort is anchored in accurate and rigorously validated annotations. Each answer is carefully reviewed for correctness and completeness, and explicitly aligned with the musical content of the corresponding score.

Strict string matching is often insufficient for capturing the semantic equivalence inherent in diverse musical terminologies, especially for generative QA, such as the relationship between ``V-I'' and ``authentic cadence''. To address this, we adopt an ensemble LLM-as-a-judge framework \citep{chen2024mllm}, utilising majority voting across three distinct models, namely, ChatGPT-5 \citep{openai2025gpt5}, Claude Sonnet 4 \citep{anthropic2025claudesonnet4}, and Gemini 2.5 Pro \citep{comanici2025gemini25pushingfrontier}, to mitigate stochasticity and bias, as illustrated in \cref{fig:data}. This methodology decouples evaluation from syntactic form, ensuring that assessment relies on semantic logic; we then verify the correctness via the human annotators. Statistical analysis confirms the reliability of this alignment, yielding a Pearson correlation \citep{pearson1896mathematical} of $0.805$ ($p < 0.001$), which confirms a significant linear association. Meanwhile, a McNemar's test \citep{mcnemar1947note} indicates no significant difference between the LLM and human evaluations ($p > 0.05$), supporting the null hypothesis of marginal homogeneity where disagreements are random rather than systematic. In addition, the inter-judge agreement across ChatGPT-5, Claude Sonnet 4, Gemini 2.5 Pro, and an independent human assessment is 83.33\%, with a 95\% bootstrap confidence interval of (79.44\%, 87.22\%), further supporting the stability of the evaluation pipeline.


\section{Benchmark Analysis}
We provide a comprehensive analysis of \BenchShort, detailing its novelty, the distribution of questions across different levels, and the characteristics of the questions. 

As shown in \cref{tab:compare}, \BenchShort evaluates LLMs and VLMs on complete musical scores, covering tasks from basic notation to advanced analysis.
Prior benchmarks offer complementary perspectives: MMMU and ZIQI-Eval prioritise multiple-choice breadth; MusicTheoryBench provides curated questions with limited score coverage; MusiXQA scales synthetic generative queries; and SSMR-Bench targets symbolic reasoning.
\BenchShort advances this landscape by unifying textual and visual modalities, enabling model trainability, and explicitly addressing homophony in full scores, a commonly overlooked aspect.
With 1,800 human-curated generative QA pairs, it balances annotation reliability with open-ended evaluation, aligning closely with current LLM and VLM research needs.

\Cref{fig:level} illustrates a balanced design in which each of the four levels accounts for 25\% of \BenchShort. More details are in \Cref{sec:question_type}.

\textbf{Level 1} emphasises performance and metadata, with onset information forming the largest proportion, complemented by smaller contributions from composer, title, tempo, and anacrusis.
\textbf{Level 2} addresses markings (14\%) and symbolic literacy, with note-level features (6\%), time values (3\%), and ornaments (2\%), and key change for modulation comprising 1\%.
\textbf{Level 3} distributes emphasis evenly across harmonic features, including chord identification (ID), cadences, tonic chords, and chord inversions (each 3\%), with progressions, suspensions (sus.), anticipations (ant.), chord spacing, and harmonic identification (ID).
\textbf{Level 4} highlights broader structural dimensions, with motif (9\%) and form (8\%) most prominent, supplemented by texture, register, tone, and other questions.

In addition, \BenchShort encompasses a wide range of composers, as shown in \Cref{fig:composer_frequency}, \Cref{sec:distribution}, spanning historical periods and stylistic traditions including the Baroque, Classical, Romantic, and twentieth-century repertoire. 
The distributions of scores by period and genre are presented in \cref{fig:period,fig:genre} from \Cref{sec:distribution}.

\begin{table*}[t]
    \setlength{\tabcolsep}{10pt}
    \centering
    \caption{Zero-shot evaluation results on \BenchShort (full), with the highest accuracy in \textbf{bold}. 
    We evaluate 12 questions per score in a single run for each model to report the accuracy for each level and overall.}
    \small
    \begin{tabular}{@{}l c c c c >{\columncolor{myblue!30}}c}
        \toprule
        \multirow{3.5}{*}{\centering\textbf{Models}} & \multicolumn{5}{c}{\textbf{\textit{\BenchLong}}} \\
        \cmidrule(lr){2-6}
        &\textbf{Level 1} & \textbf{Level 2} & \textbf{Level 3} & \textbf{Level 4} & \textbf{Overall} \\
        & $(450)$ & $(450)$ & $(450)$ & $(450)$ & $(1800)$ \\
        \midrule
        \rowcolor{myyellow!50}\multicolumn{6}{c}{\textbf{\textit{Textual QA}}} \\
        \addlinespace[0.3em]\hdashline\addlinespace[0.3em]

        Qwen3-VL-235B-A22B-Instruct &  $58.44$ 	& $33.56$& $34.00$ &	$38.89$ & $41.22$ \\
        Claude Opus 4 &  $57.11$ 	& $36.89$& $35.56$&	$35.56$&$41.28$ \\
        Claude Sonnet 4 &  $61.11$ 	& $40.67$& $35.56$&	$33.11$&$42.61$ \\
        Grok 4 & $62.00$ & $40.00$ & $31.11$ & $37.11$ & $42.61$ \\ 
        ChatGPT-5-mini & $59.11$ &  $43.56$ &	 $31.33$ &  $\textbf{40.89}$ &	$43.72$ \\
        ChatGPT-5 & $62.00$ & $50.22$ & $38.44$ & $38.44$ & $47.28$ \\
        Gemini 2.5 Pro & $\textbf{65.33}$ & $\textbf{56.00}$ & $\textbf{38.67}$ & $37.78$ & $\textbf{49.44}$ \\

        \midrule
        \rowcolor{myyellow!50}\multicolumn{6}{c}{\textbf{\textit{Visual QA}}} \\
        \addlinespace[0.3em]\hdashline\addlinespace[0.3em]
        ChatGPT-5-mini & $7.11$ &	$6.67$ &	 $8.89$	& $6.67$	& $7.33$ \\
        Grok 4 & $14.00$ & $11.11$ & $18.44$ & $21.33$ & $16.22$ \\
        Qwen3-VL-235B-A22B-Instruct &  $18.67$ 	& $15.33$& $22.44$ &	$\textbf{22.67}$ & $19.78$ \\        
        Gemini 2.5 Flash & $19.56$ &	$15.33$	& $29.56$	& $18.00$	& $20.61$ \\
        Qwen2.5-VL-72B-Instruct & $21.78$ & $18.22$ & $27.33$ & $18.89$ & $21.56$ \\  
        Claude Sonnet 4 &  $\textbf{27.11}$ 	& $16.44$& $27.33$&	$18.44$&$22.33$ \\
        Gemini 2.5 Pro & $22.00$ & $\textbf{22.44}$ & $29.11$ & $20.00$ & $23.39$ \\
        Claude Opus 4 &  $25.33$ 	& $21.78$& $\textbf{30.44}$&	$19.33$&$\textbf{24.22}$ \\
       \bottomrule
    \end{tabular}
    \label{tab:main-results}
\end{table*}

\section{Experiments}
\subsection{Experiment Settings}
\textbf{Evaluation.} 
Model outputs are evaluated through a voting process involving ChatGPT-5, Claude Sonnet 4, and Gemini 2.5 Pro, with more details provided in \Cref{sec:Inference and Evaluation Procedure,sec:appendix_prompt}. Accuracy is reported at both the individual level and the aggregate level (overall). We consider two evaluation settings: (1) zero-shot, testing models directly on the 1,800 QA pairs; and (2) fine-tuned, where models are adapted with LoRA \citep{hu2021loralowrankadaptationlarge}. We also introduce the Level-wise Success Rate (LSR), defined as the proportion of scores for which all questions are answered correctly from Level~1 through Level~$l$. Let $n$ denote the maximum level, let $l \in \{1,2,\dots,n\}$ be the level index, and let $\mathcal{S}$ denote the score set. Then, the LSR at Level $l$ is defined as
\[
LSR(l) \;=\; \frac{1}{|\mathcal{S}|} \sum_{s\in\mathcal{S}} \mathbf{1}\!\left[\bigwedge_{j=1}^{l}\;\text{AllCorrect}(s,j)\right],
\]
where $\text{AllCorrect}(s,j)$ is true if all Level-$j$ questions for score $s$ are answered correctly, and $\mathbf{1}[\cdot]$ is the indicator function. Then, we use the Wilson score interval \citep{wilson1927probable} to calculate the 95\% Confidence Interval (CI) for the LSR at Level $l$, which is given by
\[
\text{CI}(l) \;=\; 
\frac{\hat{p}_l + \tfrac{z^2}{2n_l}}{1 + \tfrac{z^2}{n_l}}
\pm
\frac{z}{1+\tfrac{z^2}{n_l}}
\sqrt{ \frac{\hat{p}_l(1-\hat{p}_l)}{n_l} + \frac{z^2}{4n_l^2} },
\]
where $\hat{p}_l$ is the LSR at Level $l$, $n_l=|\mathcal{S}|$, and $z$ is the standard normal quantile ($z=1.96$).

\noindent
\textbf{Baselines.} 
We evaluate a diverse set of models for the zero-shot evaluation, including both LLMs and VLMs. For textual QA (ABC notation), we evaluate ChatGPT-5, ChatGPT-5-mini \citep{openai2025gpt5}, Claude Opus 4 \citep{anthropic2025claudesonnet4}, Claude Sonnet 4, Deepseek-V3 \citep{deepseekai2025deepseekv3technicalreport}, Gemini 2.5 Flash \citep{comanici2025gemini25pushingfrontier}, Gemini 2.5 Pro, Grok 4 \citep{xai2025grok4}, Llama 4 Maverick \citep{meta2024llama4}, Llama 4 Scout \citep{meta2024llama4}, Qwen2.5-VL-7B-Instruct \citep{bai2025qwen25vltechnicalreport}, Qwen2.5-VL-32B-Instruct \citep{bai2025qwen25vltechnicalreport}, Qwen2.5-VL-72B-Instruct \citep{bai2025qwen25vltechnicalreport}, Qwen3-4B \citep{yang2025qwen3technicalreport}, Qwen3-32B \citep{yang2025qwen3technicalreport}, Qwen3-Max \citep{yang2025qwen3technicalreport}, and Qwen3-VL-235B-A22B-Instruct \citep{qwen3vl2025}.

For visual QA (PDF documents), we include Claude Opus 4, Claude Sonnet 4, Gemini 2.5 Flash, Gemini 2.5 Pro, ChatGPT-5-mini, Grok 4, Qwen2.5-VL-32B-Instruct, Qwen2.5-VL-72B-Instruct, and Qwen3-VL-235B-A22B-Instruct.

\noindent
\textbf{Models.} We employ Qwen3-0.6B \citep{yang2025qwen3technicalreport}, Qwen3-1.7B \citep{yang2025qwen3technicalreport}, Qwen3-4B, and Qwen2.5-VL-3B-Instruct for the fine-tuned evaluation, adapted using LoRA.

\noindent
\textbf{Data Splitting.} 
\BenchShort consists of 150 musical scores. It is divided into training, validation, and testing sets in a 6:2:2 ratio, corresponding to 90, 30, and 30 pieces, respectively. The 30-score test set is fixed and never used for training. For the fine-tuned evaluation, all test scores are identical to those used in the zero-shot evaluation.

\noindent
\textbf{Training.}
\label{sec:training_setup}
We fine-tune the models for 20 epochs on 6$\times$A800 GPUs using LoRA (rank 8). Training uses AdamW \citep{loshchilov2019decoupledweightdecayregularization} with a $5 \times 10^{-5}$ learning rate, cosine schedule, 10\% warm-up, batch size 1, and gradient accumulation of 16. For Qwen2.5-VL-3B-Instruct, we consider three types of input: PDF only, ABC notation only, and their combination (detailed in \Cref{sec:training}).

\subsection{Empirical Results}

\begin{figure*}
    \centering
    \includegraphics[width=\linewidth]{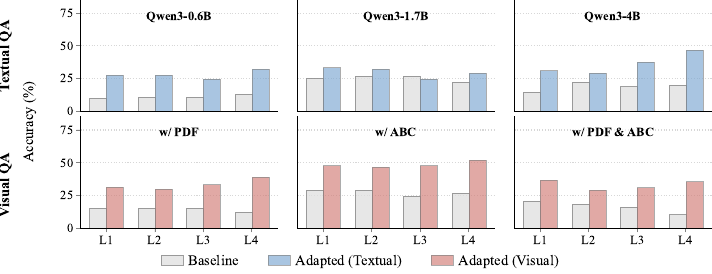}
    \caption{Performance of baseline and LoRA-adapted models on \BenchShort (testing set). Qwen2.5-VL-3B-Instruct is adapted using LoRA across the three input modalities outlined in \Cref{sec:training_setup}. L denotes Levels.}
    \label{fig:lora}
\end{figure*}

\begin{figure*}[t]
    \centering
    \begin{subfigure}[b]{0.49\textwidth}
        \centering
        \includegraphics[width=\linewidth]{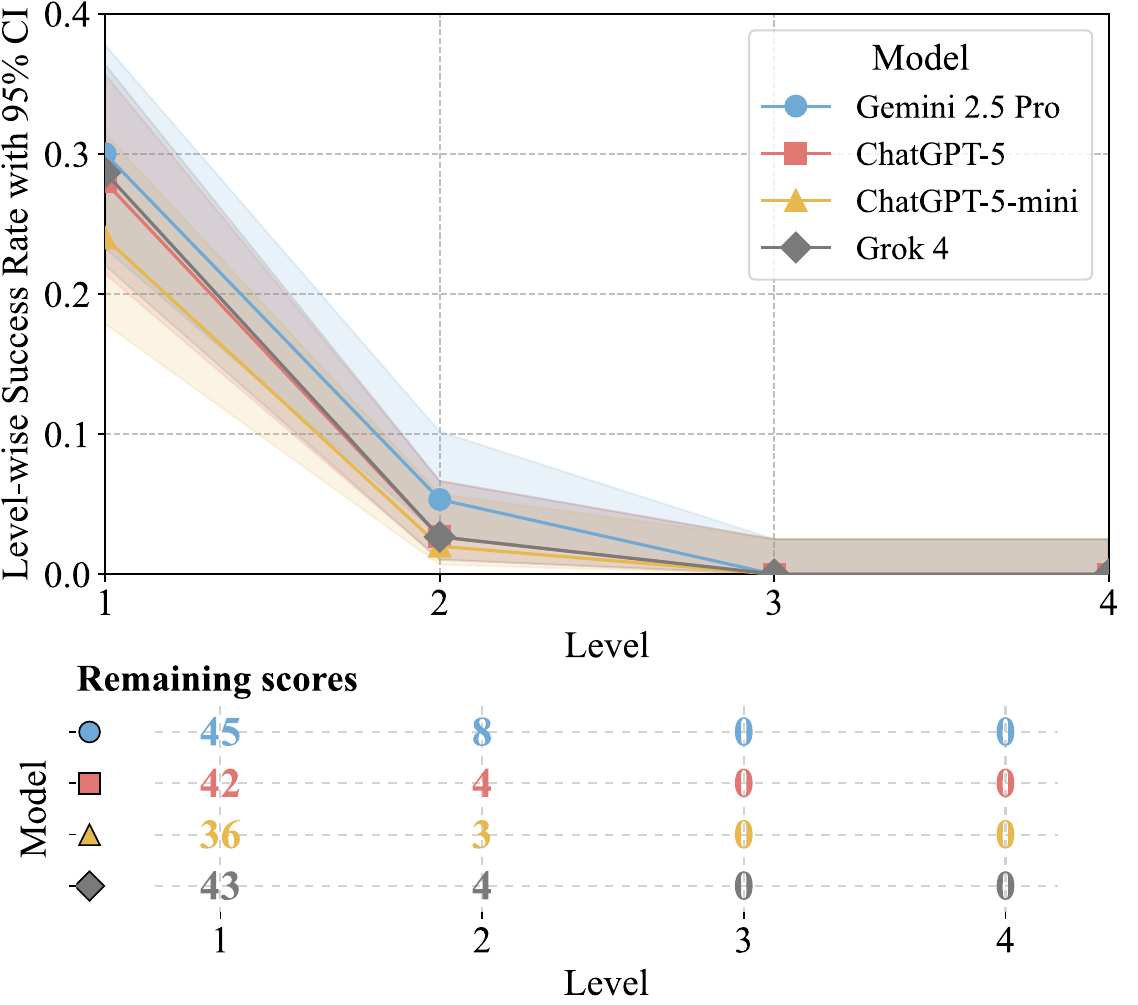}
        \caption{Textual QA.}
        \label{fig:full_abc_1800}
    \end{subfigure}
    \hfill
    \begin{subfigure}[b]{0.49\textwidth}
        \centering
        \includegraphics[width=\linewidth]{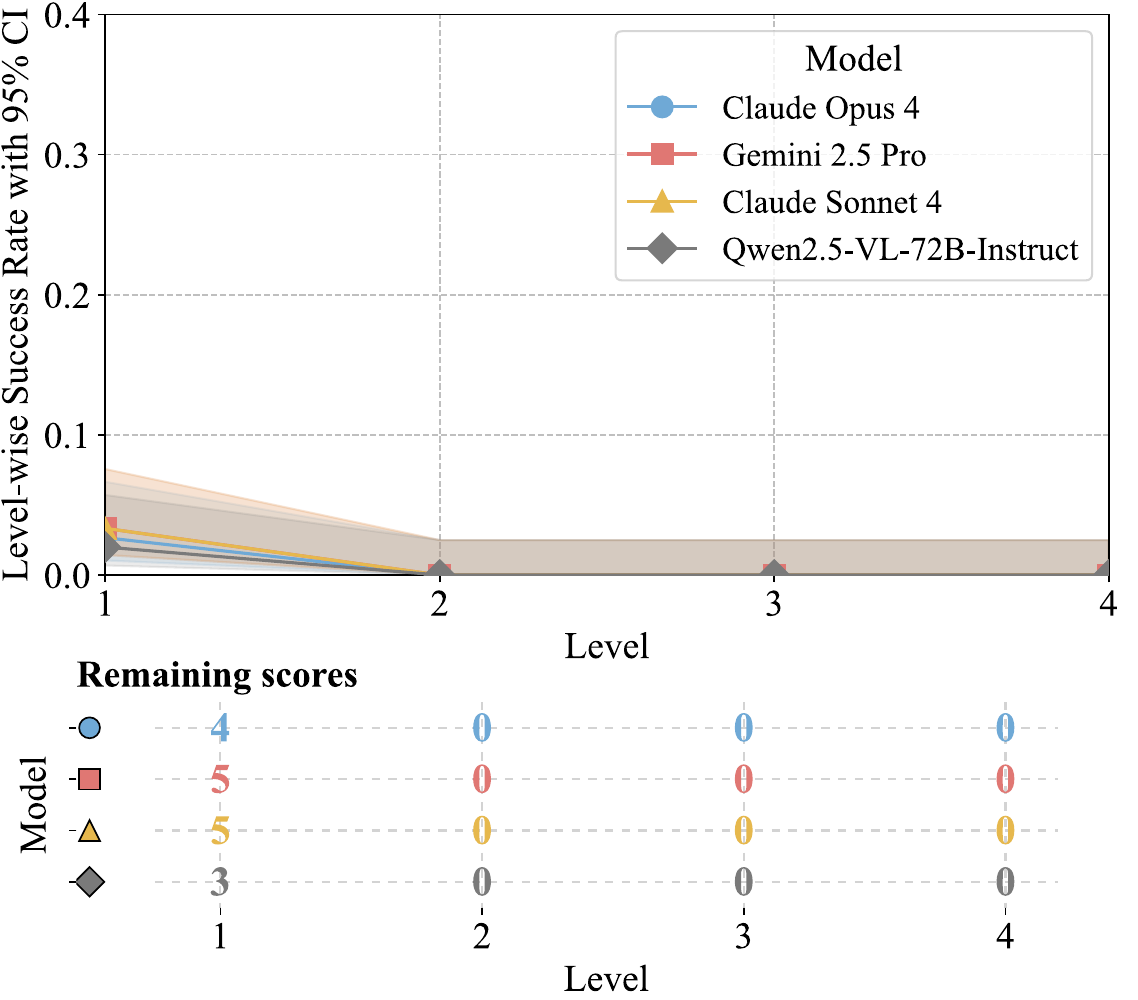}
        \caption{Visual QA.}
        \label{fig:full_pdf_1800}
    \end{subfigure}
    \caption{Level-wise Success Rate for zero-shot evaluation on MSU-Bench.}
    \label{fig:lsr_zero_shot}
\end{figure*}

\begin{table*}[t]
    \setlength{\tabcolsep}{9pt}
    \centering
    \caption{Accuracy and 95\% percentile bootstrap confidence intervals for the top-performing proprietary zero-shot models in each modality. The interval is computed as $\mathrm{CI}_{95\%} = [Q_{0.025}(\{\hat{p}^{*(b)}\}_{b=1}^{B}),\; Q_{0.975}(\{\hat{p}^{*(b)}\}_{b=1}^{B})]$, where $\hat{p}^{*(b)}$ denotes the accuracy estimate obtained from bootstrap resample $b$. All values are reported in percentage points.}
    \small
    \begin{tabular}{@{}l c c c@{}}
        \toprule
        \textbf{Modality} & \textbf{Model} & \textbf{Accuracy (\%)} & \textbf{95\% CI (\%)} \\
        \midrule
        \multirow{4}{*}{Textual QA}
        & Grok 4 & 42.61 & [40.34, 44.91] \\
        & ChatGPT-5-mini & 43.72 & [41.45, 46.02] \\
        & ChatGPT-5 & 47.28 & [44.98, 49.59] \\
        & Gemini 2.5 Pro & 49.44 & [47.14, 51.75] \\
        \midrule
        \multirow{4}{*}{Visual QA}
        & Gemini 2.5 Flash & 20.61 & [18.81, 22.54] \\
        & Claude Sonnet 4 & 22.33 & [20.47, 24.32] \\
        & Gemini 2.5 Pro & 23.39 & [21.49, 25.40] \\
        & Claude Opus 4 & 24.22 & [22.30, 26.25] \\
        \bottomrule
    \end{tabular}
    \label{tab:ci-results}
\end{table*}

\noindent
\textbf{Zero-shot Evaluation.}
In \cref{tab:main-results}, models perform substantially better on the textual QA setting than on the visual QA setting. In textual QA, Gemini 2.5 Pro achieves the best overall accuracy (49.44\%), excelling particularly at Level~1 (65.33\%) and Level~2 (56.00\%). ChatGPT-5 follows closely (47.28\%), demonstrating strong stability on higher-level questions (Levels~3--4). Notably, ChatGPT-5-mini attains the highest accuracy on Level~4 (40.89\%), suggesting an advantage in more complex reasoning despite its smaller size. 
Claude Opus~4, Claude Sonnet~4, Grok~4, and Qwen3-VL-235B-A22B-Instruct achieve comparable overall performance, with accuracies ranging from 41.22\% to 42.61\%. More results are reported in \cref{tab:more_results} in \cref{sec:more_results}. The evaluation times of models achieving more than 40\% overall accuracy are reported in \Cref{sec:time} (see \Cref{fig:time}). While models such as Gemini 2.5 Pro, ChatGPT-5, and ChatGPT-5-mini achieve the highest levels of accuracy, their evaluation times are substantially longer (more than 11 hours). Notably, Qwen3-VL-235B-A22B-Instruct requires only approximately one hour to achieve an overall accuracy of 41.22\%.
For the proprietary models in Table~\ref{tab:main-results}, we report the corresponding 95\% percentile bootstrap confidence intervals in Table~\ref{tab:ci-results}. Notably, in textual QA, the confidence intervals for Gemini~2.5~Pro and ChatGPT-5-mini do not overlap, suggesting a meaningful separation under this CI estimate.
For visual QA, overall accuracies are markedly lower (around 20\%), with the strongest model (Claude Opus~4) reaching only 24.22\%. It also achieves the highest Level~3 accuracy (30.44\%), while Claude Sonnet~4 leads at Level~1 (27.11\%) and Gemini 2.5 Pro at Level~2 (22.44\%). These results highlight a clear modality gap: ABC notation provides a more reliable representation than raw score images, where recognition and localisation errors continue to dominate performance. Some PDF-QA failure cases are provided in \Cref{sec:failure_cases}.

In \cref{fig:1v12} within \Cref{sec:1v12}, we observe that the models achieve superior performance when the 12 questions pertaining to a score are posed simultaneously, particularly for Claude Sonnet 4, Claude Opus 4, and Qwen2.5-VL-72B-Instruct. This finding suggests that presenting all questions in a batch may allow Level 1 answers to support higher-level responses in \Cref{tab:main-results}.

\noindent
\textbf{Fine-tuned Evaluation.}
\Cref{fig:lora} details the results following LoRA fine-tuning, which yields substantial gains across textual and visual QA. The textual QA results illustrate a distinct scaling trend within the Qwen3 family; accuracy improves as model size increases from 0.6B to 4B, with Qwen3-4B proving the most robust. Notably, performance improves as question complexity progresses from L1--L4, suggesting that LLMs possess strong logical inference abilities for solving harder questions using ABC notation.

For visual QA, Qwen2.5-VL-3B-Instruct exhibits a significant advantage when using ABC inputs only, clearly outperforming Qwen3-4B, likely because ABC reduces the visual noise inherently present in PDF files.

\noindent
\textbf{LSR Analysis.}
\Cref{fig:full_abc_1800,fig:full_pdf_1800} show the LSR from Levels~1--4 across all 150 scores in \BenchShort. In textual QA (\Cref{fig:full_abc_1800}), models perform moderately at Level~1 (25--35\%), with Gemini~2.5~Pro slightly ahead of ChatGPT-5 and Grok~4, but drop below 10\% by Level~2 and nearly vanish by Level~3. 
Visual QA (\Cref{fig:full_pdf_1800}) is worse: models start at 5--10\% on Level~1 and collapse almost entirely by Level~2. 
The ``remaining scores'' count reinforces this gap, with about 41.5 scores surviving past Level~1 in textual QA versus only 4.25 in visual QA. These results highlight that while models can solve isolated questions, sustaining correctness across levels is extremely difficult, underscoring LSR’s diagnostic strictness; LSR for LoRA-adapted models is also reported in \Cref{sec:lsr}.

\noindent
\textbf{Massive Multitask Language Understanding (MMLU).} 
As shown in \Cref{tab:mmlu} from \cref{sec:mmlu}, the models adapted with LoRA exhibit minimal forgetting, with performance remaining close to that of their base versions tested on MMLU \citep{hendryckstest2021}. 
These results indicate that LoRA adaptation effectively preserves the models' general knowledge while enhancing their capabilities in musical score understanding and reasoning.

\begin{table}[htbp]
\centering
\small
\caption{Performance evaluation using title-only input. L1--L4 denote Levels 1--4 of MSU-Bench.}
\vspace{-5pt}
\setlength{\tabcolsep}{5pt}
\renewcommand{\arraystretch}{1.2}
\begin{tabular}{lcccc}
\toprule
Models & L1 & L2 & L3 & L4 \\
\midrule
Qwen3-0.6B & 17.7\% & 14.4\% & 11.1\% & 10.0\% \\
Qwen3-1.7B & 23.3\% & 22.2\% & 24.4\% & 24.4\% \\
Qwen3-4B & 24.4\% & 30.0\% & 26.6\% & 23.3\% \\
Qwen2.5-VL-3B & 15.5\% & 22.2\% & 20.0\% & 15.6\% \\
\bottomrule
\end{tabular}
\label{tab:noise}
\vspace{-4pt}
\end{table}

\noindent
\textbf{Analysis of Score Content Reliance.}
Following \citet{kumar2025mmauprochallengingcomprehensivebenchmark} and \citet{zang2025reallylisteningboostingperceptual}, we test whether models use local musical cues or rely on title-based guessing. In \Cref{tab:noise}, title-only input yields weak performance across Levels 1--4, suggesting that answers cannot be reliably inferred from titles alone and that models instead depend on musical information provided in ABC notation or PDF scores when such input is available.


\section{Conclusion}
We introduce \BenchLong (\BenchShort), a benchmark for evaluating complete musical score reasoning in LLMs and VLMs across textual (ABC) and visual (PDF) modalities. Our experiments reveal that while current models struggle with multi-level comprehension, particularly in visual QA, the use of ABC notation can mitigate part of this issue by providing precise bar information.

\newpage
\section*{Limitations}
While ABC notation effectively mitigates visual processing errors, the substantial performance gap observed between textual and visual modalities highlights the limitations of current models in interpreting raw sheet music without symbolic aid. In this work, ABC should therefore be interpreted as a structured upper-bound modality for symbolic reasoning rather than as evidence of fully solved visual score understanding. Computational constraints regarding token limits necessitated the exclusion of certain longer scores during the multimodal fine-tuning experiments. In addition, the scope of MSU-Bench is primarily centred on the Western art music canon, spanning the Baroque to the early twentieth century. We seek to broaden this musical scope to encompass non-Western, contemporary, and more diverse notational practices in future work.

\section*{Acknowledgements}
This work is supported by the Advanced Discipline Construction Project of Beijing Universities, the Special Programme of National Natural Science Foundation of China (Grant No. T2341003), and the Major Programme of National Social Science Fund of China (Grant No. 21ZD19).

\bibliography{custom}

\appendix
\crefalias{section}{appendix}
\section{Appendix: List of Musical Scores}
\label{sec:score}

\begin{enumerate}[leftmargin=2.2em,labelsep=0.6em]
    \item Cello Suite No.1 BWV 1007 - 1. Prélude
    \item Solfeggietto in C minor
    \item Toccata and Fugue in D minor BWV 565
    \item Fugue in G Minor BWV 542
    \item Fugue I in C major BWV 846
    \item Fugue in D minor BWV 948
    \item Fugue in G Minor BWV 578
    \item Prelude I in C major BWV 846
    \item Sonate No. 16 1st Movement
    \item Piano Sonata No. 5 in C Minor Op.10 No.1
    \item Sonata in G Op.14 No.2 Movement 1
    \item Piano Sonata in A major Op.2 No.2
    \item Piano Sonata No. 3 in C Major Op. 2 No. 3
    \item Sonata No. 23 Op. 57 Appassionata
    \item Sonata Op.31 No.17 in D minor Tempest
    \item Piano Sonata No. 17 in D minor Op. 31 \\ No. 2
    \item Piano Sonata No.18 in E flat major \\ Op.31 No.3
    \item Sonate No.8 Op.13 Pathétique 3 Rondo. \\ Allegro Sonata No.8
    \item Les Nuits d'été
    \item Symphonie fantastique, H 48
    \item Polovtsian Dances
    \item Hungarian Dance No. 5
    \item Rhapsody Op. 79 No. 2
    \item Waltz Op.39 No.3
    \item Intermezzo in E flat major Op.117 No.1
    \item B minor Rhapsody 1 Op. 79
    \item Ballade Op.118 No.3
    \item Intermezzo Op. 116 No. 2
    \item Intermezzo Op. 118 No. 2 A Major
    \item Violin Concerto in E minor Op.64
    \item Lullaby Op.49 No.4
    \item Waltz in A Major Op.39 No.15
    \item Fantaisie-Impromptu in C$\sharp$ Minor
    \item Nocturne Op. 9 No.1
    \item Nocturne-No. 20 in C Sharp Minor
    \item Ballade no.1 in G minor Op.23
    \item Sonata No.2 Op.35 1st Movement
    \item Ballade No.3 in A flat major Op.47
    \item Ballade No.4 in F minor Op
    \item Prélude Opus 28 No. 4 in E Minor
    \item Waltz in A Minor
    \item Nocturne Op. 27, No. 2
    \item Suite Bergamasque
    \item La fille aux cheveux de lin
    \item Reverie
    \item Clair de lune
    \item Premier Trio
    \item Syrinx
    \item Sonate pour Violoncelle et Piano
    \item Symphony No. 9 New World II, Largo
    \item Symphony No. 9 New World:IV, Allegro con fuoco
    \item Humoresque No.7
    \item Holberg Suite Op.40 I.Praeludium
    \item Wedding Day at Troldhaugen
    \item Anitras Dance Piano solo
    \item Dance Op. 12 No. 4
    \item Sailors Song Op.68 No.1
    \item Waltz Op. 12 No. 2
    \item Butterfly Sommerfugl Op. 43 No. 1
    \item Piano Concerto in A minor Op.16
    \item In the Hall of the Mountain King
    \item Lyric Pieces Op.47 Grieg
    \item Lyric Pieces Op. 54 No. 4
    \item Morning Mood from Peer Gynt Suite No. 1
    \item Sonata in E Minor, Hob. XVI: 34 (I: Presto)
    \item String quartet - Op.76, No.5, in D major
    \item Cello Concerto C Major Movement 1
    \item Piano Sonata in F Major HOB.XVI/23
    \item Haydn Sonata Hob. XVI37 Mov. 1 D Major
    \item String Quartet Op.64 No.3
    \item Piano Concerto in D major
    \item Die Schöpfung Mit Würd’ und Hoheit angetan
    \item Piano Sonata in E minor HOB. XVI/34
    \item Sonata in C minor HOB/XVI:20
    \item String Quartet in C major (“Emperor”) Op. 76 No. 3
    \item Die Fledermaus Grunfeld Op. 56 Konzertparaphrase
    \item Radetzky March
    \item Pizzicato Polka Arranged for Piano Solo
    \item The Blue Danube Accordion Solo
    \item Tratsch-Polka Op.214
    \item Strauss Die Fledermaus Op.362 Overture
    \item Hungarian Rhapsody No. 2
    \item Etude S.136 No.4
    \item Trois Etudes de Concert No. 3
    \item Der Müller Und Der Bach. D795, S.5652
    \item Hungarian Rhapsody No. 6
    \item Etude S.136 No.5
    \item Etude S.136 No.9
    \item William Tell Overture Finale
    \item Romance S.169
    \item Grandes études de Paganini, S.141: No. 6
    \item S. 1413 in G$\sharp$ Minor, La Campanella
    \item S.541 No.3 in A$\flat$ Major
    \item Symphony No.10 - I. \\ Adagio Complete Score
    \item Song Without Words Op.85 No.3
    \item Song without Words Op. 38 No.6
    \item Song Without Words Op.30 No.5
    \item Melodie Op.4 No.2 in C minor
    \item Songs without words Op.30 No.1
    \item Songs Without Words Op.19 No.6
    \item Songs Without Words Op.62
    \item Wedding March
    \item Songs Without Words Op.19 No.4
    \item Song Without Words Op.19b No.1
    \item Songs Without Words Op.19 No.3
    \item Piano Sonata No.1 K.279
    \item Sonata No. 5 1st Movement K.283
    \item Sonata No. 7 1st Movement K. 309
    \item Piano Sonata No.8 in A minor K.310300d
    \item Piano Sonata No. 8 in D Major, \\ K. 311 (284c): I. Allegro con spirito
    \item Piano Sonata No.18 in D major K
    \item Piano Concerto No.23 in A major K.488
    \item Mozart Rondo Alla Turca
    \item Piano Sonata No. 16 - Allegro
    \item Sonata No.11 in A major K.331
    \item Pictures at an Exhibition: No.2, Il \\ vecchio castello
    \item Pictures at an Exhibition 13 8. Catacombae
    \item Pictures at an Exhibition 14 Cum mortuis in lingua mortua
    \item Pictures at an Exhibition Movement 15 (No.9)
    \item Pictures at an Exhibition 16 10
    \item Pictures at an Exhibition-Gnomus (The Gnome) \& Promenade
    \item Pictures at an Exhibition
    \item Strauss Die Fledermaus Op.410 Overture
    \item Piano Concerto in G major - II
    \item Gaspard de la Nuit, No.2, ``Le Gibet'' 
    \item Gaspard de la Nuit, No. 1, ``Ondine''
    \item Flight of the Bumblebee Piano
    \item Concerto No.1 in A minor
    \item Sans The Cuckoo in the Depths of the Woods
    \item Sans - Fossils Transcribed for Piano
    \item 2nd Piano Concerto 1st Movement Piano solo
    \item Introduction and Rondo Capriccioso Op.28
    \item Le Cygne The Swan Easy Piano by Free MusicKey
    \item Allegro Appassionato Cello Piano
    \item Piano Sonata D.784 - 1st movement
    \item Impromptu in C minor No.1 Op.90
    \item Impromptu No. 3 Op. 90 D 899 G$\flat$ Major Transcription
    \item Impromptu No.3 Op.90 D 899 G Majeur Transcription de Liszt
    \item Piano Sonata No.19 in C minor
    \item Sonata Op.42
    \item Ave Maria
    \item Winterreise D.911 No.1 - Gute Nacht
    \item Winterreise D.911 No.5 - Der Lindenbaum
    \item Die Forelle D. 550 Op. 32
    \item Winterreise D.911 No.24 - Der Leiermann
    \item Schwanengesang D.957 No.4
    \item Waltz Op. 18 no. 6 in B minor
    \item Piano Sonata No.2 in G
    \item Vers La Flamme Op.72
    \item Etude Opus 8 No. 12 in D Minor
\end{enumerate}

\section{Appendix: Sample Questions}
\label{sec:sample_question}
Using \textit{Solfeggietto in C minor} by C. P. E. Bach as an illustrative example, the following section presents sample questions from each of the four levels in \BenchShort. 
In total, the dataset comprises 1,800 QA pairs drawn from the 150 complete musical scores, with 450 questions allocated to each level. The structure of these questions is exemplified below. 
Here, \textbf{Q} denotes the question and \textbf{A} denotes the reference answer.
\\ \\
\textbf{Level 1}
\begin{itemize}
    \item \textbf{Q1:} Is the piece written with an anacrusis (upbeat bar)?\\
          \textbf{A1:} No.
    \item \textbf{Q2:} What is the tempo in beats per minute?\\
          \textbf{A2:} 150 bpm.
    \item \textbf{Q3:} What is the initial key?\\
          \textbf{A3:} C minor.
\end{itemize}

\noindent
\textbf{Level 2}
\begin{itemize}
    \item \textbf{Q1:} In bar 1, what is the dynamic marking?\\
          \textbf{A1:} f.
    \item \textbf{Q2:} In bar 11, is there an accidental, and what is it?\\
          \textbf{A2:} F$\sharp$4, A$\natural$4.
    \item \textbf{Q3:} In bar 25, what is the ornament on note D2?\\
          \textbf{A3:} Passing note, neighbour note.
\end{itemize}

\noindent
\textbf{Level 3}
\begin{itemize}
    \item \textbf{Q1:} In bar 33, what is the chord progression? (use I, II, III, IV, V, VI, VII)\\
          \textbf{A1:} I--V7.
    \item \textbf{Q2:} In bar 27, what scale degree is the first chord? (use I, II, III, IV, V, VI, VII)\\
          \textbf{A2:} I.
    \item \textbf{Q3:} In bar 12, is there a dominant chord?\\
          \textbf{A3:} No.
\end{itemize}

\noindent
\textbf{Level 4}
\begin{itemize}
    \item \textbf{Q1:} What is the predominant rhythm of the piece?\\
          \textbf{A1:} Semiquaver.
    \item \textbf{Q2:} In bar 1, in which register is the melody?\\
          \textbf{A2:} Middle register.
    \item \textbf{Q3:} What are the main features of the motif?\\
          \textbf{A3:} Third interval.
\end{itemize}

\section{Appendix: Instructions Given to Annotators}
\label{sec:Instructions Given To Annotators}

As the annotators are experienced, we provide instructions only on how to format reference answers, as follows:

\begin{enumerate}
    \item For enquiries regarding ornaments, identify the category of the ornament, such as ``appoggiatura'' or ``mordent'', rather than listing the specific auxiliary pitches;
    \item Utilise Scientific Pitch Notation (SPN) for all pitch identifications. Ensure that the octave number is included immediately after the note name (such as C4 rather than C);
    \item For harmonic analysis, represent chords using Roman numerals to denote scale degrees and qualities (such as I, V7, or iv) rather than using macro-analytical symbols or jazz lead sheet notation;
    \item When identifying keys, format the response with the capitalised note name followed by the mode in lowercase, formatted as ``[Note] major'' or ``[Note] minor'';
    \item If a single question elicits multiple distinct answers, separate each item with a comma to facilitate accurate verification.
\end{enumerate}

\section{Appendix: Recruitment and Payment}
\label{sec:Recruitment And Payment}

We recruited ten doctoral candidates specialising in music from leading conservatoires in China. All participants began their formal musical training in early childhood, and consequently, each individual possessed approximately two decades of musical experience. Participation in this study was entirely voluntary and uncompensated. The decision to forgo monetary compensation was deemed adequate given the status of the participants as junior experts intrinsically motivated to contribute to academic discourse, rather than crowdsourced workers requiring financial incentives. This volunteer-based model is consistent with the norms of academic participation for this specific demographic in the region and adheres strictly to the ethical guidelines approved by the host institution.

\section{Appendix: Detailed Breakdown of Question Types for Levels 1--4}
\label{sec:question_type}
In this section, we provide a comprehensive breakdown of the question types encompassed within each level of \BenchShort, as illustrated in \Cref{fig:level}.

\subsection{Level 1}
\textbf{Onset Information.} The questions ask about onset information that appears at the very beginning of a musical score, including the composer and title, the initial key, clef, and time signature, the presence of an anacrusis, the instruments involved, as well as the opening tempo, metronome marking, and expression indications. 
Collectively, these elements establish the basic identity, notation, and performance instructions that frame how the piece is read and interpreted from the outset. \\

\subsection{Level 2}
\textbf{Notation \& Note.} Level 2 questions differ from Level 1 questions in both scope and depth. 
Whereas Level 1 focuses on onset information that is immediately visible at the start of a score, such as composer, title, initial key, time signature, clef, instrumentation, tempo, and expression indications, Level 2 shifts attention to localised details within the body of the music. 
These questions aim to examine specific bars for note-level features (highest or lowest pitch, presence of accidentals, shortest or longest note and rest values), performance instructions (dynamics, articulation, ornaments, tempo changes, expression markings), and structural signs (repeat signs, clef or key changes, modulation).

\subsection{Level 3}
\textbf{Harmony \& Chord.} Level 3 questions delve into mid-level musical structures, focusing on harmonic and chordal analysis within specific bars.
Level 3 progresses from the recognition of individual symbols to the analytical reasoning required for understanding harmonic structure. 
Rather than focusing on isolated surface features, such as a dynamic marking or an accidental within a single bar, these questions require the interpretation of tonal function. 
LLMs or VLMs are expected to identify chords by scale degree using Roman numerals, determine inversions and spacing, recognise cadences including perfect, imperfect, and interrupted, and distinguish non-chord tones such as suspensions (sus.) and anticipations (ant.). 
In addition, this level addresses harmonic progressions extending across multiple bars, examines whether passages commence or conclude on the tonic, and considers the treatment of chord tones that are omitted or repeated.

\subsection{Level 4}
\textbf{Texture \& Form.} Level 4 extends beyond the recognition of notation and harmony to encompass piece-specific understanding and large-scale structural analysis. 
The questions require the identification of a work’s genre and performance medium, including its orchestration or ensemble, as well as its formal design, such as principal and secondary themes, transitions, and main sections. 
They also address thematic materials by asking where a motif first appears, its defining features in terms of rhythm, register, or instrumental part, and the ways in which it is developed or ornamented. 
Further areas of focus include prevailing textural and accompanimental conventions, the predominant tempo and rhythmic character of the work, and, on occasion, the number of movements. 
In the case of instrument-specific repertoire, idiomatic features such as bowings are considered. 
Overall, Level 4 demands the synthesis of information across extended spans of music in order to characterise style, form, texture, and thematic organisation, thereby moving towards holistic musical analysis rather than bar-by-bar observation.

\section{Appendix: Distribution of Composers, Periods, and Genres in \BenchShort}
\label{sec:distribution}

We provide a detailed account of the composer coverage, historical periods, and genre distribution of the musical scores included in \BenchShort. The purpose of this analysis is to show that the benchmark is not narrowly tailored to a single stylistic tradition, but instead reflects a broad and representative spectrum of Western art music commonly encountered in formal music education and musicological analysis.

\subsection{Composer Distribution}

\begin{figure*}[t]
    \centering
    \includegraphics[width=1\linewidth]{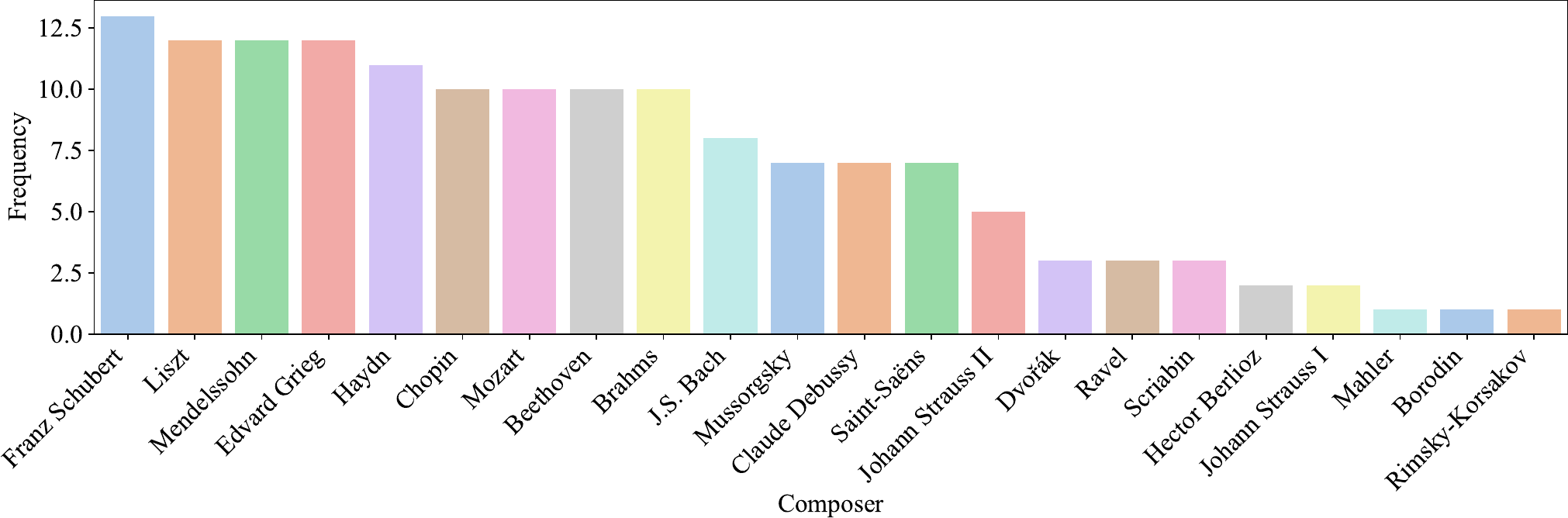}
    \caption{Frequency distribution of composers represented in \BenchShort. The histogram illustrates the number of pieces per composer, with Franz Schubert, Liszt, Mendelssohn, and Edvard Grieg appearing most frequently, while representation gradually decreases for others.}
    \label{fig:composer_frequency}
\end{figure*}

\BenchShort encompasses works by a diverse set of composers who occupy central positions in the Western canon, as shown in \cref{fig:composer_frequency}. The selected scores include representative works by Johann Sebastian Bach, Ludwig van Beethoven, Fr\'ed\'eric Chopin, Johannes Brahms, Claude Debussy, Modest Mussorgsky, Hector Berlioz, Robert Schumann, Franz Liszt, and Pyotr Ilyich Tchaikovsky, among others. This selection strategy ensures coverage of composers associated with distinct compositional languages, formal conventions, and notational practices. As a result, the benchmark exposes models to a wide range of stylistic fingerprints, from contrapuntal writing and functional harmony to extended chromaticism and colouristic textures.

\subsection{Historical Periods}
As shown in \cref{fig:period}, the musical scores in \BenchShort span a broad historical range, covering the Baroque, Classical, Romantic, and twentieth-century repertoire. The distribution is weighted toward Romantic music, while still maintaining meaningful representation from earlier and later traditions. This temporal spread exposes models to contrasting compositional idioms, including contrapuntal writing, formal balance, chromatic expansion, and more fluid twentieth-century harmonic and textural practices. Such diversity is important for evaluating whether models can generalise musical reasoning beyond a single historical style.

\subsection{Genre and Formal Categories}
\begin{figure*}[htbp]
    \centering
    \begin{subfigure}[b]{0.48\textwidth}
        \centering
        \includegraphics[width=\textwidth]{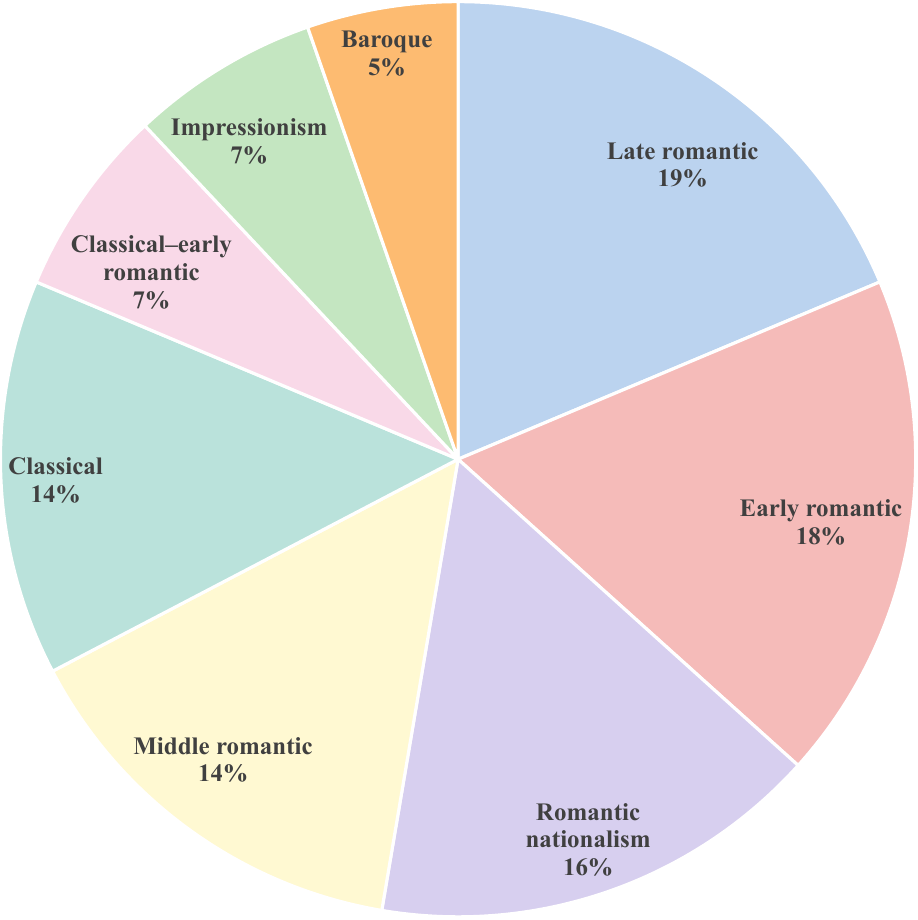}
        \caption{Period}
        \label{fig:period}
    \end{subfigure}
    \hfill
    \begin{subfigure}[b]{0.486\textwidth}
        \centering
        \includegraphics[width=\textwidth]{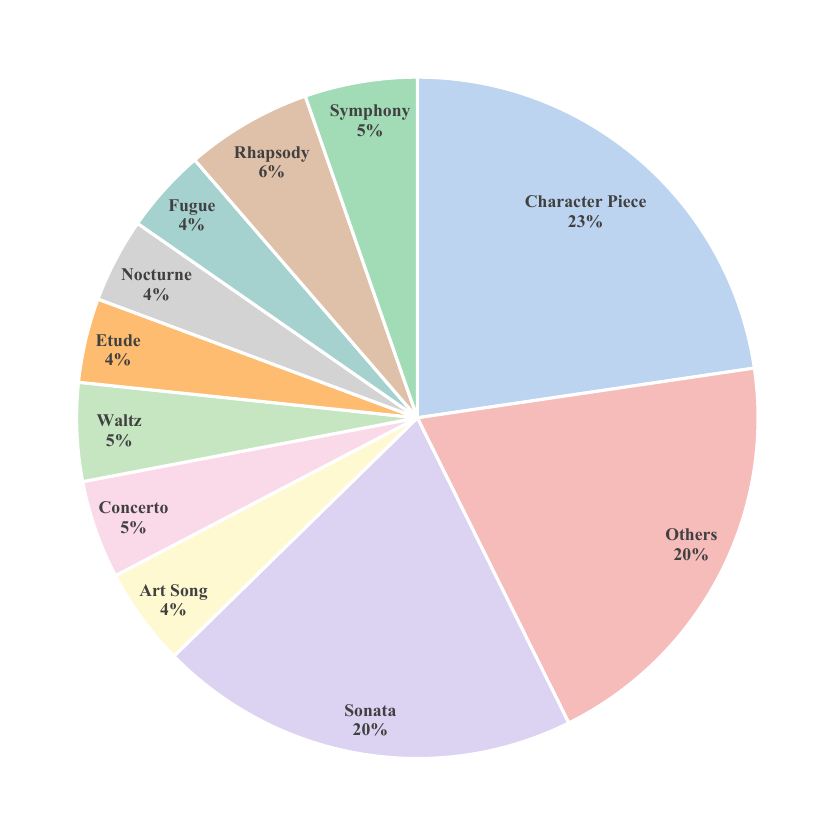}
        \caption{Genre}
        \label{fig:genre}
    \end{subfigure}
    \caption{Distribution of musical periods and genres in \BenchShort. (a) shows the historical periods of the selected scores, ranging from Baroque (5\%) to Impressionism (7\%) and various stages of Romanticism, with Late Romantic (19\%) and Early Romantic (18\%) being most prominent. (b) presents the genre distribution, where character pieces and sonatas each account for 23\%, followed by others (20\%), rhapsodies (6\%), concertos and symphonies (5\% each), waltzes (5\%), and fugues, nocturnes, etudes, and art songs, each comprising 4\%.}
\end{figure*}

As shown in \cref{fig:genre}, \BenchShort covers a broad range of genres and formal categories, including sonatas, character pieces, fugues, waltzes, nocturnes, etudes, rhapsodies, symphonies, concertos, art songs, and others. These genres differ not only in surface style but also in their structural and textural demands: contrapuntal works require attention to voice leading and imitation, whereas sonatas, symphonic movements, and concertos more often involve large-scale formal reasoning. This distribution reflects the coexistence of extended forms and shorter self-contained works, allowing the benchmark to assess not only local notational understanding but also broader reasoning about form, thematic organisation, and texture across diverse compositional contexts.

\section{Appendix: Inference and Evaluation}
\label{sec:Inference and Evaluation Procedure}
Inference and evaluation are conducted entirely on a local GPU environment (A800). We use vLLM \citep{10.1145/3600006.3613165} as the inference backend and adopt the unified data loading and prompt templating mechanisms provided by LLaMA-Factory \citep{zheng-etal-2024-llamafactory} to ensure a consistent inference pipeline. The evaluated models include both unmodified base models and parameter-efficiently fine-tuned variants using LoRA. Whether a LoRA adapter is loaded during inference is explicitly controlled by configuration settings, ensuring fair comparability across different model variants.

During inference, the model generates answers autoregressively given the musical score images and the corresponding question text. Random sampling is used for decoding, including temperature sampling as well as top-p and top-k sampling. In our experiments, the decoding parameters are consistently set to temperature = 0.95, top-p = 0.7, and top-k = 50.

During the LLM-as-a-judge evaluation, for each sample, the judgement models take a predicted answer and a reference answer as input and produce a binary decision (1 for correct, 0 for incorrect). To ensure determinism, a fixed prompt template in \Cref{sec:appendix_prompt} is used, and the decoding temperature is set to 0, allowing only a single judgement token to be generated.

Across all evaluated samples, the inter-judge agreement among ChatGPT-5, Claude Sonnet 4, Gemini 2.5 Pro, and an independent human assessment is 83.33\%, with a 95\% bootstrap confidence interval of (79.44\%, 87.22\%). This high unanimity rate indicates substantial consistency across judges and supports the reliability of the automatic evaluation pipeline.


\section{Appendix: Prompt Details}
\label{sec:appendix_prompt}

This appendix presents the exact prompt templates used throughout our inference and evaluation pipeline. The prompts are designed to support both general music-theoretic questions and score-localisation-based questions, while also providing a standardised instruction format for the LLM-as-a-judge setting. The same template is applied across models to ensure consistency in prompting and evaluation. For localisation-based questions, the model is instructed to identify the relevant musical position before producing an answer, thereby reducing unsupported guessing and mitigating hallucinations in position-sensitive tasks. For the judgement setting, the prompt enforces a strict binary decision based on semantic equivalence between the predicted answer and the reference answer.

\subsection{Inference User Prompt}

\begin{lstlisting}
Here is a piece of music in {score_modality}:
{score}

Question:
{Question}
\end{lstlisting}

\subsection{Inference System Prompt}

\begin{lstlisting}
You are a professional assistant specialised in music theory and musical scores.
You can answer both conceptual music questions and score-based questions.
If a question requires locating a specific position in a musical score
(such as a chapter, movement, section, bar, measure, beat, staff, or voice),
you must first identify that position in the score and then read the required information from it.

Rules:
- Determine whether the question is score-localisation-based or not.
- For score-localisation questions, do NOT guess or generalise.
- For non-localisation questions, answer using standard music theory knowledge.
- If the required information cannot be determined with certainty, output "Unknown".
- Do NOT provide unnecessary explanations unless explicitly asked.

Your output should be accurate, concise, and directly answer the question.
\end{lstlisting}

\newpage
\subsection{LLM-as-a-Judge Prompt}
\begin{lstlisting}
You are an expert in music theory. Your task is to evaluate the correctness of a Predicted Answer based on a Reference Answer.

Reference Answer: "{reference}" Predicted Answer: "{prediction}"

Instruction:

Relying on your music theory knowledge, compare the semantic meaning of the Predicted Answer to the Reference Answer.

If the Predicted Answer matches the Reference Answer, output 1.

If the Predicted Answer contradicts or fails to match the Reference Answer, output 0.

Do not output any text or explanation. Output only the single digit.

Decision: \end{lstlisting}

\begin{figure*}[h]
    \centering
    \includegraphics[width=0.8\linewidth]{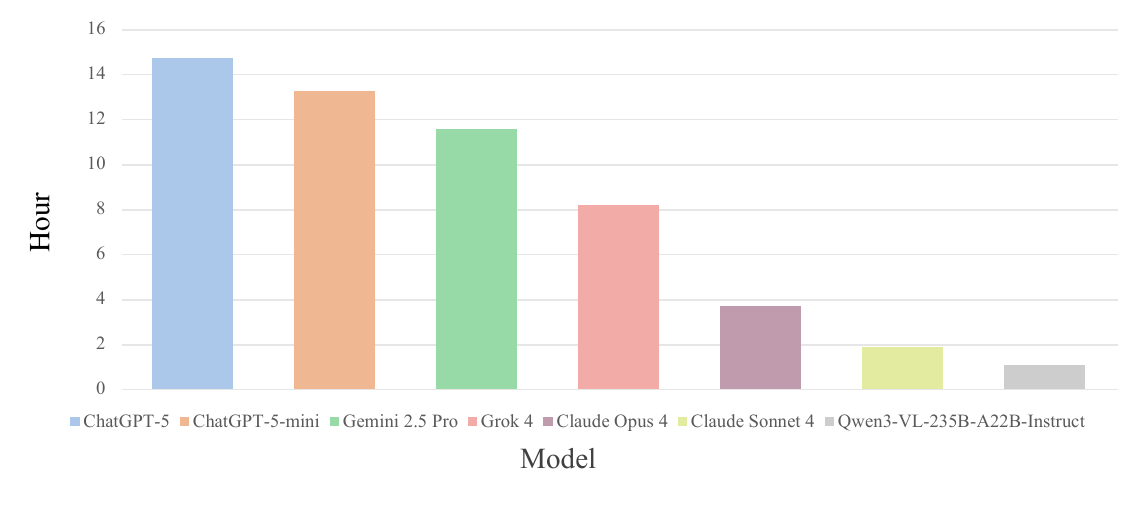}
    \caption{The inference time for models exceeding 40\% overall accuracy.}
    \label{fig:time}
\end{figure*}

\begin{table*}[t]
    \setlength{\tabcolsep}{10pt}
    \centering
    \caption{More zero-shot evaluation results on \BenchShort (full), with the highest accuracy in \textbf{bold}. 
    We evaluate 12 questions per score in a single run for each model to report the accuracy for each level and overall.}
    \small
    \begin{tabular}{@{}l c c c c >{\columncolor{myblue!30}}c}
        \toprule
        \multirow{3.5}{*}{\centering\textbf{Models}} & \multicolumn{5}{c}{\textbf{\textit{\BenchLong}}} \\
        \cmidrule(lr){2-6}
        &\textbf{Level 1} & \textbf{Level 2} & \textbf{Level 3} & \textbf{Level 4} & \textbf{Overall} \\
        & $(450)$ & $(450)$ & $(450)$ & $(450)$ & $(1800)$ \\
        \midrule
        \rowcolor{myyellow!50}\multicolumn{6}{c}{\textbf{\textit{Textual QA}}} \\
        \addlinespace[0.3em]\hdashline\addlinespace[0.3em]

        Qwen3-4B        & $20.00$ & $10.00$ & $8.67$ & $13.11$ & $12.94$\\
        Qwen2.5-VL-7B-Instruct & $32.00$ & $9.11$ & $17.56$ & $13.33$ & $18.00$ \\ 
        Qwen2.5-VL-72B-Instruct & $34.44$ & $18.00$ & $18.89$ & $12.67$ & $21.00$ \\  
        Llama 4 Scout &  $48.44$ 	& $25.78$ & $26.89$ &	$26.44$ & $31.89$ \\
        Qwen2.5-VL-32B-Instruct        & $50.67$ & $20.22$ & $26.22$ & $37.56$ & $33.67$\\
        Gemini 2.5 Flash & $50.22$ &	$31.11$	& $30.67$	& $24.89$	& $34.22$ \\
        Qwen3-Next-80B-A3B-Instruct & $\mathbf{57.11}$ & $23.11$ & $25.33$ & $34.00$ & $34.89$ \\
        Deepseek-V3 & $52.89$ & $\mathbf{32.67}$ & $30.22$ & $29.56$ & $36.33$ \\
        Llama 4 Maverick &  $52.67$ 	& $31.56$& $28.44$&	$33.56$ & $36.56$ \\
        Qwen3-Max &  $54.67$ 	& $31.56$& $\mathbf{31.78}$&	$\mathbf{40.67}$&$\mathbf{39.67}$ \\
       \bottomrule
    \end{tabular}
    \label{tab:more_results}
\end{table*}

\section{Appendix: Different Training Settings}
\label{sec:training}
To investigate the effect of different input modalities on model adaptation, we design three distinct fine-tuning strategies for Qwen2.5-VL-3B-Instruct.

\noindent
\textbf{PDF.}  
In this setting, we treat PDF sheet music as the visual input modality. The model receives images rendered from PDF pages, and both the visual encoder and the language model are updated during training. This setting evaluates the model’s ability to extract structural and symbolic information directly from visual sheet-music representations.

\noindent
\textbf{ABC notation.}  
Here, we replace PDF images with ABC notation as the only training input. Since this modality does not require visual parsing, we freeze the visual encoder to reduce computational overhead and update only the language model and LoRA adapters. This strategy evaluates whether ABC notation alone is sufficient for enabling VLMs to learn music-theoretical patterns.

\noindent
\textbf{PDF + ABC notation.}  
In the multimodal setting, we provide both PDF images and ABC notation for each score. Both the visual encoder and the language model are fine-tuned. The objective is to examine whether complementary information from visual and symbolic modalities produces better performance than unimodal training. By integrating structural cues from PDF files with explicit symbolic tokens from ABC, this approach is expected to enhance robustness and generalisation across diverse tasks. However, the combination of both modalities constrains the maximum number of tokens available for training. The affected long scores are removed from the training pool only before the train/validation split is finalised; the fixed 30-score test set is never used for training and is unaffected by these exclusions. The excluded training scores are:

\begin{enumerate}
    \item \textit{Piano Sonata No.~5 in C Minor, Op.~10 No.~1}
    \item \textit{Sonate pour Violoncelle et Piano}
    \item \textit{Piano Concerto in A Minor, Op.~16}
    \item \textit{Hungarian Rhapsody No.~2}
    \item \textit{Ballade No.~1 in G Minor, Op.~23}
    \item \textit{Ballade No.~4 in F Minor, Op.~52}
    \item \textit{Sonata, Op.~42}
    \item \textit{Concerto No.~1 in A Minor}
\end{enumerate}

\begin{figure*}[ht]
    \centering
    \includegraphics[width=0.95\linewidth]{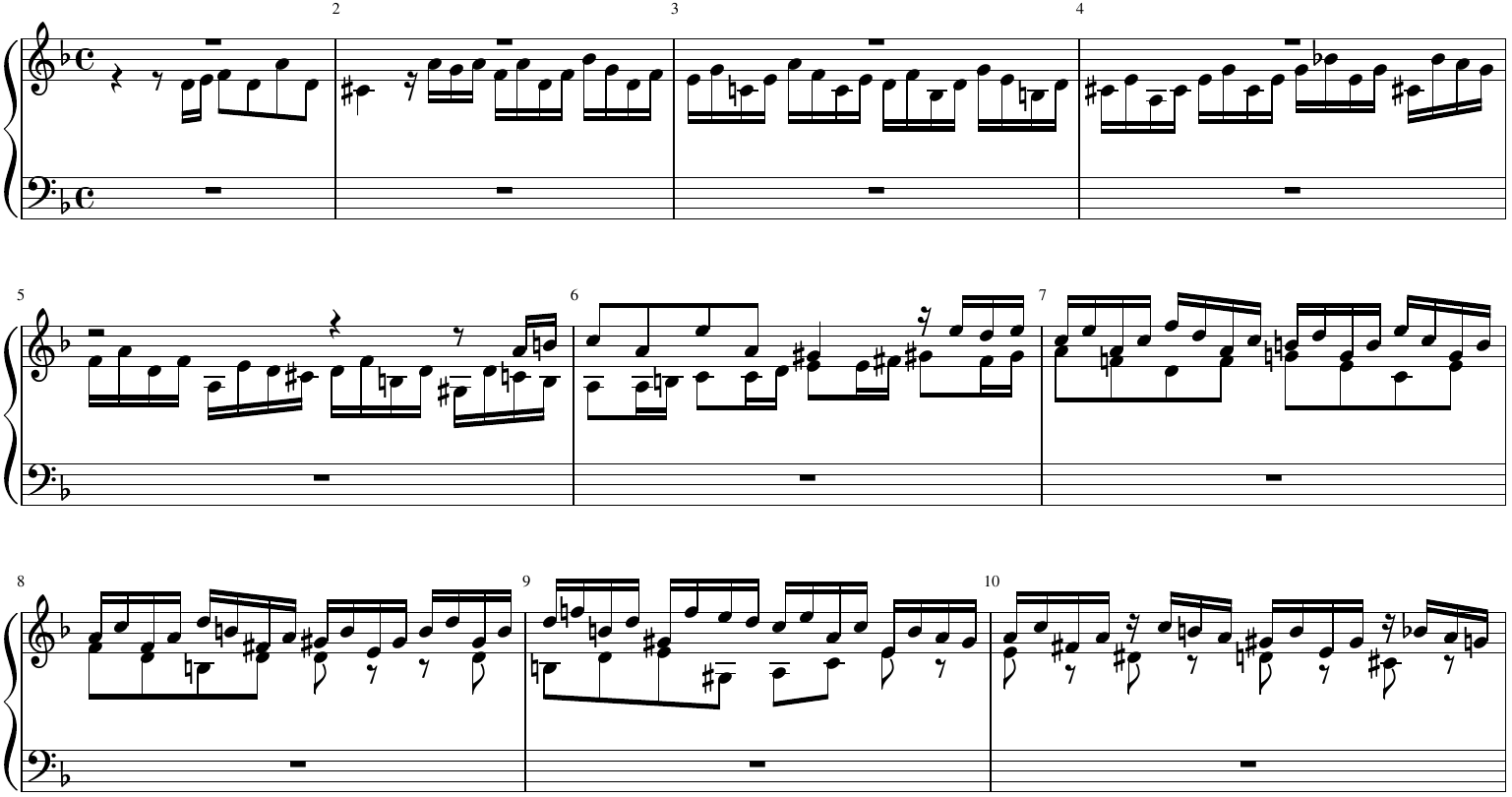}
    \caption{Representative score excerpt for the Level 2 and Level 4 PDF-QA failure cases.}
    \label{fig:failure_case_l2}
\end{figure*}

\begin{figure*}[ht]
    \centering
    \includegraphics[width=0.95\linewidth]{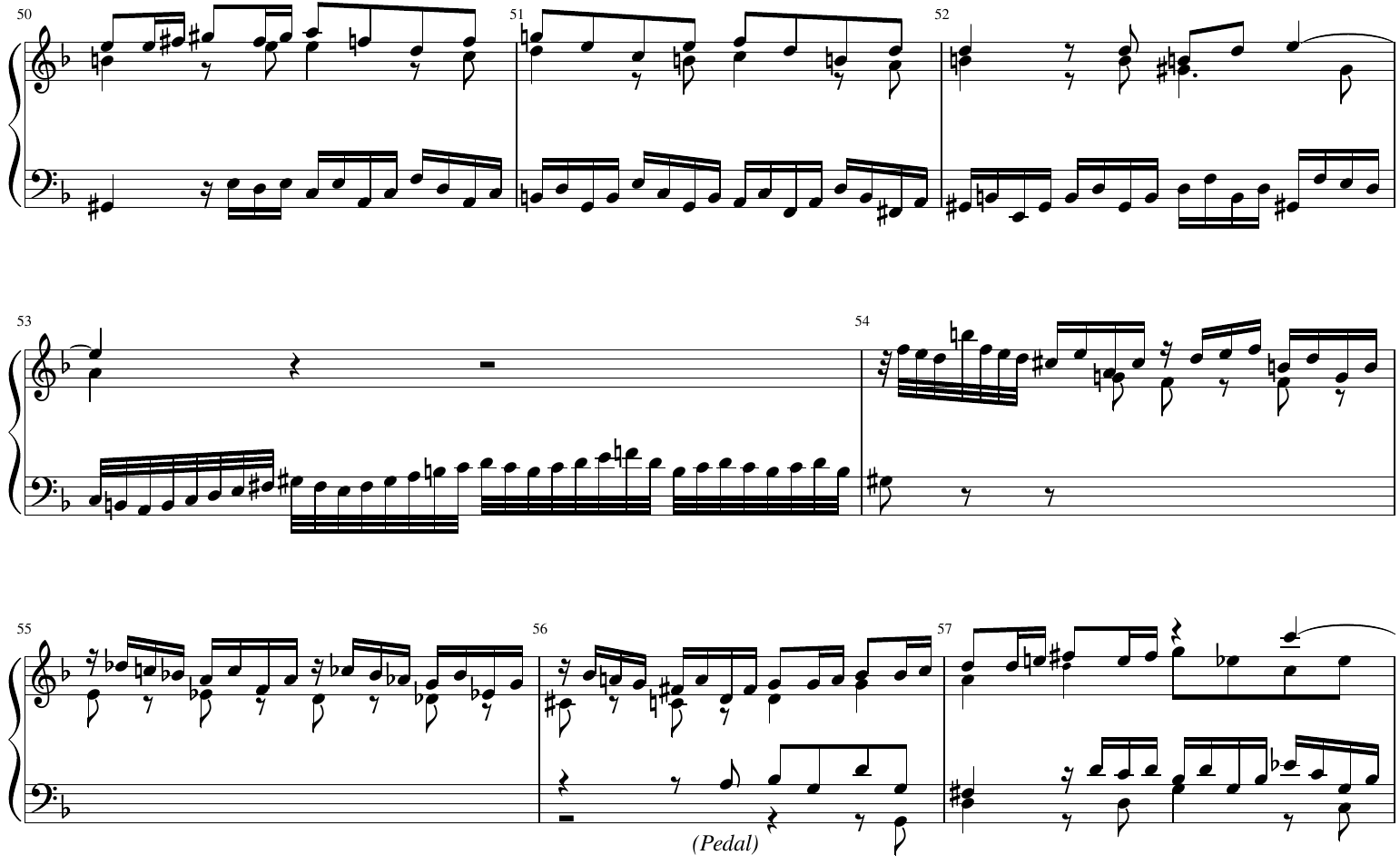}
    \caption{Representative score excerpt for the Level 3 PDF-QA failure case.}
    \label{fig:failure_case_l3}
\end{figure*}

\section{Appendix: More results for Baselines}
\label{sec:more_results}
In \cref{tab:more_results}, we report additional zero-shot baseline results on the full \BenchShort benchmark to provide a broader comparison beyond the strongest models discussed in the main text. Models such as Qwen3-Max and Llama~4 Maverick remain below 40\% overall accuracy, indicating that strong general-purpose language capabilities do not directly translate into robust musical score understanding. Among the open-source models evaluated, Qwen3-VL-235B-A22B-Instruct demonstrates the strongest overall performance, exceeding the text-only Qwen3-Max by about 4\%. This result suggests that large-scale multimodal modelling provides a measurable advantage even when performance is still far from saturation. In contrast, smaller models such as Qwen3-4B and Qwen2.5-VL-7B-Instruct perform considerably worse, thereby highlighting the limitations of lightweight architectures in zero-shot musicological reasoning tasks. The level-wise breakdown further shows that performance gains are not confined to a single difficulty level, but instead reflect more stable improvements across the benchmark as model capacity increases.

\section{Appendix: Inference Time}
\label{sec:time}
We report the wall-clock inference time required by models achieving more than 40\% overall accuracy in the zero-shot setting, corresponding to the results in \cref{tab:main-results} and \cref{fig:time}. The evaluation is conducted on the full MSU-Bench.

Overall, a clear trade-off between performance and computational cost is observed. Large proprietary models, including Gemini~2.5~Pro, ChatGPT-5, and ChatGPT-5-mini, attain the highest accuracies but require substantially longer evaluation times, typically exceeding eleven hours for a complete run. This latency arises from a combination of model size, inference overhead, and limited parallelism in sequential generative question answering.

In contrast, Qwen3-VL-235B-A22B-Instruct achieves a competitive overall accuracy of 41.22\% while requiring only approximately one hour of evaluation time. This result highlights the efficiency advantages of certain open-source architectures, which deliver favourable accuracy--efficiency trade-offs under identical evaluation protocols.

These findings suggest that inference time constitutes a practical constraint that should be considered alongside raw accuracy. Efficient models may offer a more viable solution for rapid development and comparative analysis, even when absolute performance differences are relatively small.

\begin{figure*}[ht]
    \centering
    \includegraphics[width=\linewidth]{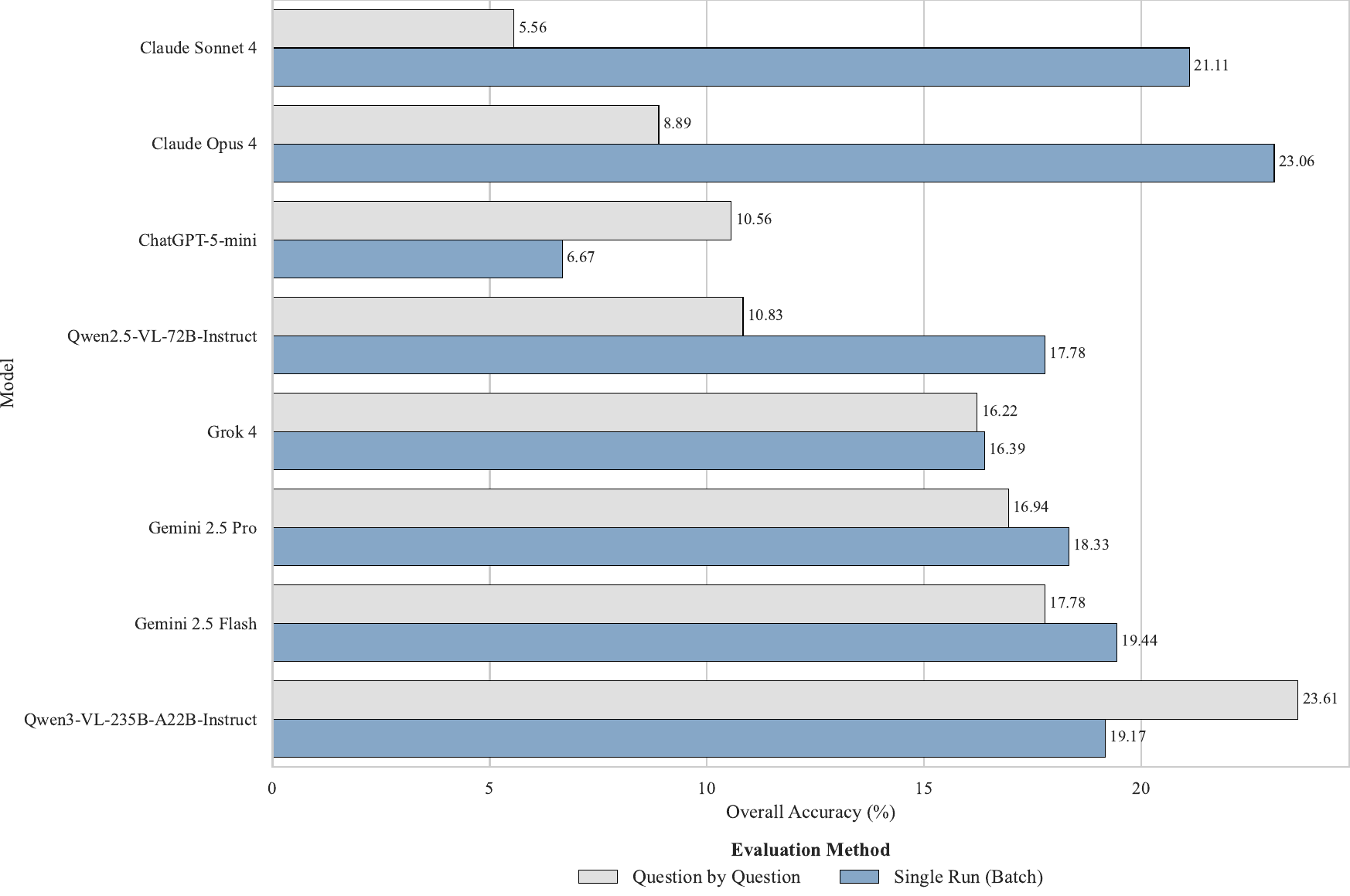}
    \caption{Overall accuracy comparison under the Question-by-Question and Single-Run (Batch) evaluation for VLMs. The horizontal axis reports overall accuracy in percentage, while each bar corresponds to a specific model.}
    \label{fig:1v12}
\end{figure*}

\section{Appendix: Representative PDF-QA Failure Cases}
\label{sec:failure_cases}
To complement the aggregated visual-QA results, we present three representative PDF-QA failure cases on Johann Sebastian Bach’s Fugue in D minor, BWV 948, generated by Claude Opus 4 (Figures~\ref{fig:failure_case_l2} and \ref{fig:failure_case_l3}). These examples highlight common error patterns in full-score reasoning, particularly incorrect bar-level grounding and hallucinated local details. In the following examples, \textbf{Q} denotes the question, \textbf{P} the model prediction, and \textbf{GT} the ground truth.

\subsection{Level 2 failure case}
\noindent\textbf{Q:} In bar 15, what is the highest pitch?

\noindent\textbf{P:} Looking at bar 15 on page 2, the right hand (treble clef) begins with some notes featuring sharps and then moves through a melodic line. The highest notes in the treble staff in bar 15 appear to reach up to approximately D5 (the D above middle C, sitting just below the top of the treble staff). There are notes on the upper portion of the staff, but nothing ventures above the staff lines into ledger-line territory. The highest pitch in bar 15 is D5.

\noindent\textbf{GT:} A4.

\noindent\textbf{Observation:} The model misreads the local pitch content in the target bar and overestimates the upper note range, indicating weak note-level grounding even when the queried bar is explicitly specified and the relevant notes are visually accessible.

\subsection{Level 3 failure case}
\noindent\textbf{Q:} In bar 54, is there a perfect cadence?

\noindent\textbf{P:} It doesn't render the actual notes clearly enough for me to read the precise pitches and harmonies. What I can see from the layout is that bar 54 sits in the middle of a passage on page 6, but I can't reliably determine the specific chords to confirm whether there's a V–I (perfect/authentic) cadence at that point.

\noindent\textbf{GT:} No.

\noindent\textbf{Observation:} Instead of resolving the harmonic relation in the target bar, the model fails to ground the local notation strongly enough to make a cadence judgement and abstains from answering.

\subsection{Level 4 failure case}
\noindent\textbf{Q:} In which bar does the melodic motif first appear?

\noindent\textbf{P:} The fugue subject (the main melodic motif) first appears in bar 2 in the right hand — it's the opening solo entry that begins with a rest on beat 1, then launches into the characteristic rising figure followed by running sixteenth notes. This is the subject that gets developed throughout the fugue. Bar 1 is essentially empty (a full-bar rest), so the motif's first appearance is in bar 2. Subsequent entries of the subject appear in various bars as the fugue unfolds (e.g., the bass voice enters later around bar 11–12 with the answer).

\noindent\textbf{GT:} Bar 1.

\noindent\textbf{Observation:} This error reflects incorrect bar anchoring at the opening of the piece: the model imposes an inferred rest-based reading and shifts the first motif entry from bar 1 to bar 2.

\begin{figure*}[h]
    \centering
    \begin{subfigure}[b]{0.49\textwidth}
        \centering
        \includegraphics[width=\linewidth]{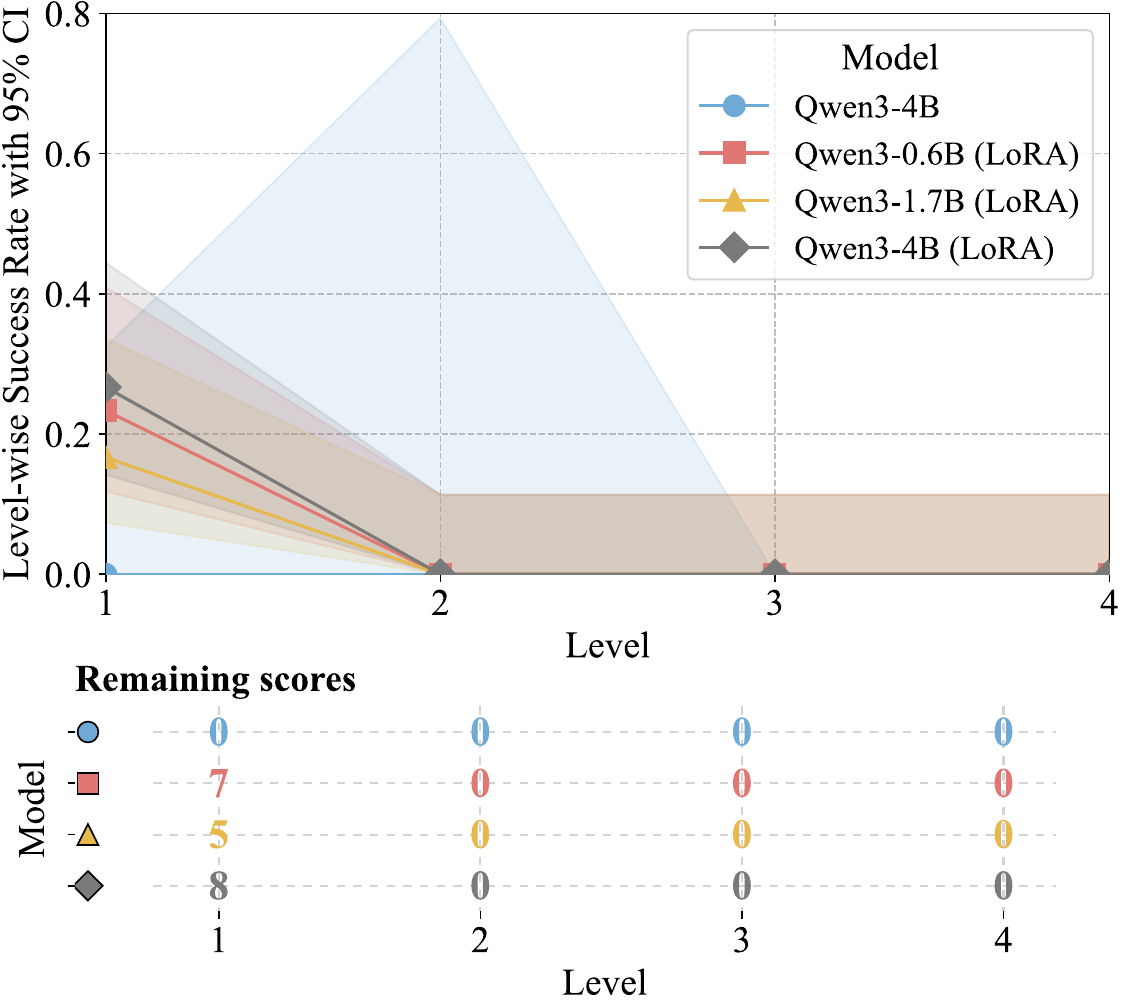}
        \caption{LLMs. Results for models trained solely on ABC notation without other modalities. The LoRA-finetuned models demonstrate a discernible advantage over the base Qwen3-4B model.}
        \label{fig:abc_lora}
    \end{subfigure}
    \hfill
    \begin{subfigure}[b]{0.49\textwidth}
        \centering
        \includegraphics[width=\linewidth]{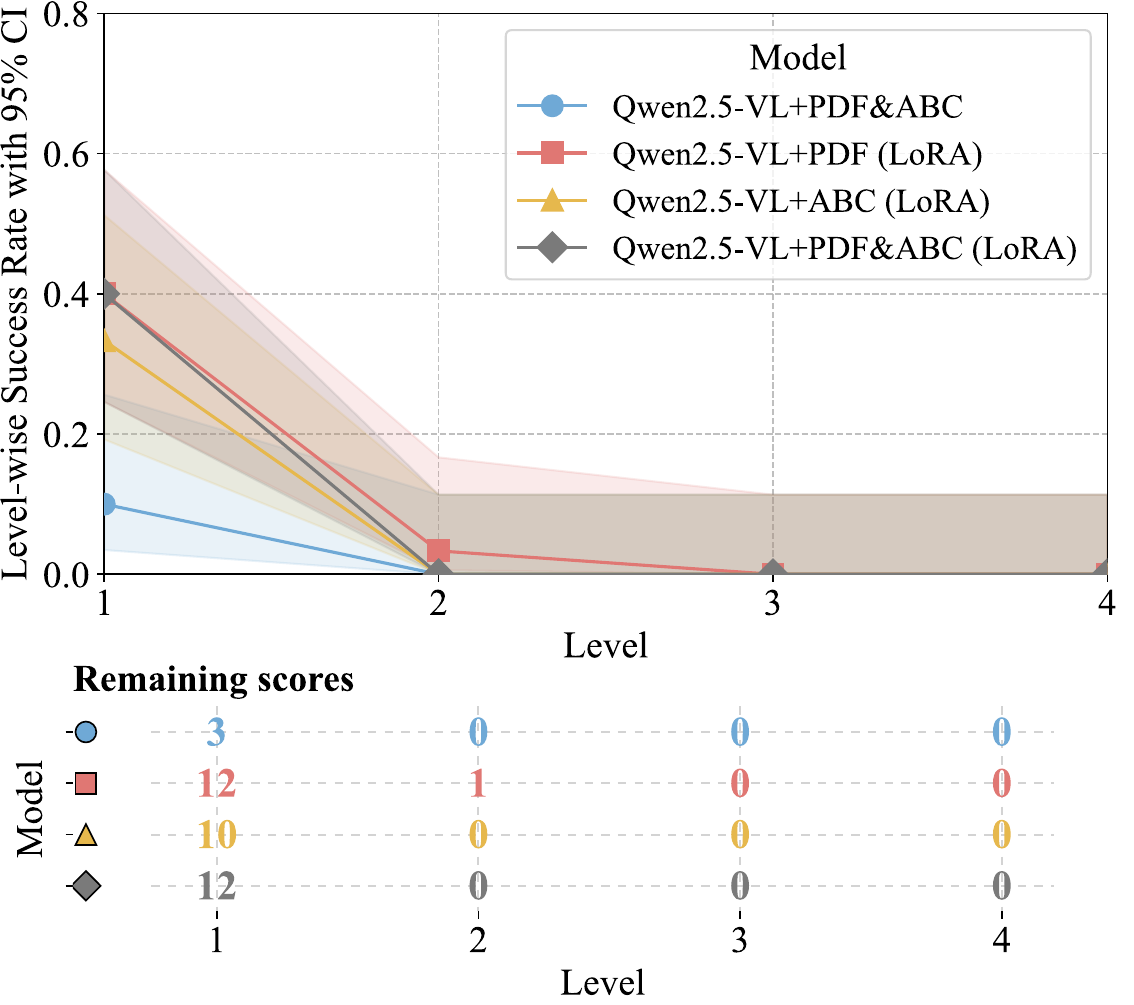}
        \caption{VLMs. We compare VLMs trained on PDF, ABC, or both modalities against the baseline Qwen2.5-VL-3B-Instruct, which utilises combined inputs.}
        \label{fig:pdf_lora}
    \end{subfigure}
    \caption{Level-wise Success Rate for Models Adapted Using LoRA. }
    \label{fig:lsr_lora}
\end{figure*}

\section{Appendix: One vs Batched Questions}
\label{sec:1v12}

In \Cref{fig:1v12}, we compare two evaluation settings for VLMs on \BenchShort: asking each question individually and asking all 12 questions associated with a score in a single batch. The results reveal substantial variation across models, with batched prompting generally leading to higher overall accuracy. This pattern suggests that jointly presenting the questions allows models to better exploit shared score-level context, especially when lower-level observations can support answers to higher-level questions. At the same time, the magnitude of the gain differs across models, indicating that some systems benefit more than others from cross-question contextualisation within a single inference run.

\section{Appendix: Analysis of LSR for Models Adapted Using LoRA}
\label{sec:lsr}

We also report LSR for models adapted with LoRA, complementing the zero-shot LSR analysis in \Cref{fig:full_pdf_1800}. LSR provides a stricter view of model behaviour than per-question accuracy, because a score is counted as successful at Level~$l$ only if all questions from Level~1 through Level~$l$ are answered correctly. As a result, it directly captures whether a model can maintain consistency across progressively more demanding stages of musical understanding rather than solving isolated questions in isolation.

Consistent with the zero-shot results, LSR decreases monotonically as the level increases, reflecting the growing difficulty of sustaining correctness across successive levels of musical comprehension. Even when a model performs reasonably well on individual questions, small errors at earlier levels accumulate and sharply reduce the number of scores that remain fully correct at later levels. This makes LSR especially informative for evaluating end-to-end score understanding.

However, LoRA-adapted models exhibit markedly improved retention compared to their base counterparts, as shown in \Cref{fig:abc_lora,fig:pdf_lora}. For LLMs, adapted models maintain non-negligible success rates compared to the base model (Qwen3-4B), indicating that lightweight supervised adaptation helps preserve correctness across multiple levels rather than only boosting isolated responses. Improvements are even more pronounced in VLMs. While zero-shot visual models fail almost entirely after Level~1, LoRA adaptation enables models such as Qwen2.5-VL-3B-Instruct to preserve a meaningful number of answerable scores at Level~1, especially when PDF and ABC inputs are combined.

Taken together, these results indicate that LoRA adaptation not only improves per-question accuracy but also enhances cross-level consistency, mitigating the compounding error effects revealed by LSR. Nevertheless, the sharp decline at higher levels persists across all settings, underscoring that robust, end-to-end musical reasoning over complete scores remains an open challenge even after adaptation.

\section{Appendix: MMLU}
\label{sec:mmlu}

\begingroup
\setlength{\tabcolsep}{15pt}
\begin{table*}[t]
    \centering
    \caption{Evaluation of models conducted before and after LoRA on MMLU. Qwen2.5-VL-3B-Instruct is adapted using LoRA across the three input modalities described in \Cref{sec:training_setup}.}
    \resizebox{\textwidth}{!}{%
    \begin{tabular}{@{\extracolsep{\fill}} l S S S S S@{}}
        \toprule
        \centering\textbf{Models}                             & \textbf{STEM} & \textbf{Humanities} & \textbf{Social Sciences} & \textbf{Other Subjects} \\
        \midrule
        Qwen3-4B        & 72.63 & 81.44 & 63.21 & 74.61 \\
        \quad\quad\quad\quad 
        w/ LoRA  & 74.09\textsuperscript{\textcolor{magenta}{(+01.46)}} 
                     & 81.54\textsuperscript{\textcolor{magenta}{(+00.10)}} 
                     & 63.51\textsuperscript{\textcolor{magenta}{(+00.30)}} 
                     & 75.11\textsuperscript{\textcolor{magenta}{(+00.50)}} \\
        \addlinespace[0.3em]\hdashline\addlinespace[0.3em]
        Qwen2.5-VL-3B-Instruct        & 60.60 & 75.63 & 58.72 & 69.65 \\
        \quad\quad\quad\quad
        w/ PDF  & 60.90\textsuperscript{\textcolor{magenta}{(+0.30)}} 
                     & 75.66\textsuperscript{\textcolor{magenta}{(+00.03)}} 
                     & 58.45\textsuperscript{\textcolor{magenta}{(-00.27)}} 
                     & 69.80\textsuperscript{\textcolor{magenta}{(+00.15)}} \\
        \quad\quad\quad\quad 
        w/ ABC      & 60.47\textsuperscript{\textcolor{magenta}{(-00.13)}}  
                     & 75.79\textsuperscript{\textcolor{magenta}{(+00.16)}}  
                     & 58.13\textsuperscript{\textcolor{magenta}{(-00.59)}} 
                     & 69.62\textsuperscript{\textcolor{magenta}{(-00.03)}}  \\
        \quad\quad\quad\quad 
        w/ PDF\&ABC & 60.50\textsuperscript{\textcolor{magenta}{(-00.10)}}  
                     & 75.85\textsuperscript{\textcolor{magenta}{(+00.22)}}  
                     & 58.28\textsuperscript{\textcolor{magenta}{(-00.44)}} 
                     & 69.65\textsuperscript{\textcolor{magenta}{(-00.00)}} \\
        \bottomrule
    \end{tabular}
    } 
    \label{tab:mmlu}
\end{table*}
\endgroup

We evaluate the models adapted with LoRA and those without adaptation on MMLU in order to assess whether task-specific adaptation to \BenchShort leads to catastrophic forgetting in general-domain knowledge. MMLU covers 57 subjects spanning Science, Technology, Engineering, and Mathematics (STEM), as well as the humanities, social sciences, and other professional domains, and therefore provides a broad reference benchmark for retained reasoning ability.

The results indicate that LoRA adaptation does not cause substantial degradation on MMLU. For the Qwen3-4B model, the application of LoRA yields consistent improvements across all reported categories, including STEM, Humanities, Social Sciences, and Other Subjects. This suggests that adaptation to the musical benchmark can in some cases improve general instruction-following or transfer behaviour rather than weakening it. In contrast, the performance of Qwen2.5-VL-3B-Instruct remains largely stable following adaptation with different input modalities, including PDF and ABC notation. The observed deviations are small, indicating that modality-specific adaptation does not materially erode the model’s general knowledge.

Overall, these findings suggest that the proposed fine-tuning setup is relatively robust with respect to forgetting: models can be adapted to musical score understanding while largely preserving their competence on a broad external benchmark. This is important in practice, as it indicates that gains on \BenchShort are not simply obtained by sacrificing more general reasoning ability.

\section{Appendix: Use of LLMs}
\label{sec:use_of_llm}
In preparing this manuscript, we employ LLMs solely as an auxiliary tool for academic writing. Their use is restricted to \textit{linguistic refinement}, including polishing grammar, improving clarity and fluency, 
and adjusting the structure and formatting of text and tables. We do not rely on LLMs for generating research ideas, designing methodologies, conducting experiments, performing data analysis, or interpreting results. 
All conceptual contributions, experimental designs, and substantive findings reported in this work are entirely our own. All final decisions regarding content, wording, and presentation are made by the authors.

\end{document}